\documentclass[sigplan, 10pt]{acmart}
\usepackage{savesym}
\usepackage{amsmath}
\usepackage{amsthm}
\usepackage{subfig}
\usepackage{graphicx}
\usepackage{xspace}
\usepackage{setspace}
\usepackage{stackrel}
\usepackage{verbatim}
\usepackage{xcolor}
\usepackage[linesnumbered,ruled,noline]{algorithm2e}

\SetCommentSty{mycommfont}
\usepackage{color}
\usepackage{ifthen}
\usepackage{comment}
\usepackage{graphicx}
\usepackage{enumerate}
\usepackage{mathrsfs}
\usepackage{picture}
\usepackage{overpic}
\usepackage{multirow}
\usepackage{booktabs}
\usepackage[all]{xy}
\usepackage{pbox}
\usepackage[noend]{algpseudocode}
\usepackage{tikz}
\savesymbol{mathdollar}
\usepackage{MnSymbol}
\restoresymbol{abc}{mathdollar}

\SetKwFor{For}{for (}{) $\lbrace$}{$\rbrace$}

\begin{document}


\newcommand{\cubic}{\begin{tikzpicture}[scale=0.15]\pgfmathsetmacro{\cubex}{1}\pgfmathsetmacro{\cubey}{1}\pgfmathsetmacro{\cubez}{1}\draw (0,0,0) -- ++(-\cubex,0,0) -- ++(0,-\cubey,0) -- ++(\cubex,0,0) -- cycle;\draw (0,0,0) -- ++(0,0,-\cubez) -- ++(0,-\cubey,0) -- ++(0,0,\cubez) -- cycle;\draw (0,0,0) -- ++(-\cubex,0,0) -- ++(0,0,-\cubez) -- ++(\cubex,0,0) -- cycle;\end{tikzpicture}}

\newcommand{\cub}{\begin{tikzpicture}[scale=5]\pgfmathsetmacro{\cubex}{1}\pgfmathsetmacro{\cubey}{1}\pgfmathsetmacro{\cubez}{1}\draw (0,0,0) -- ++(-\cubex,0,0) -- ++(0,-\cubey,0) -- ++(\cubex,0,0) -- cycle;\draw (0,0,0) -- ++(0,0,-\cubez) -- ++(0,-\cubey,0) -- ++(0,0,\cubez) -- cycle;\draw (0,0,0) -- ++(-\cubex,0,0) -- ++(0,0,-\cubez) -- ++(\cubex,0,0) -- cycle;\end{tikzpicture}}

\newcommand{\spaceincomplex}{\;}
\newcommand{\algorithmicname}{\textbf{}}
\newcommand{\paper}{\section}
\newcommand{\code}{\emph}
\newcommand{\mycomment}{}
\newcommand{\point}{\subsection}
\newcommand{\bigo}{O}
\renewcommand{\emph}{\textsl}
\newcommand{\tabincell}[2]{\begin{tabular}{@{}#1@{}}#2\end{tabular}}


\title{Temporal Vectorization for Stencils}

\author{Liang Yuan}
\affiliation{%
  \institution{SKL of Computer Architecture, ICT, CAS}
}
\email{yuanliang@ict.ac.cn}

\author{Hang Cao}
\affiliation{%
  \institution{SKL of Computer Architecture, ICT, CAS}
}

\author{Yunquan Zhang}
\affiliation{%
  \institution{SKL of Computer Architecture, ICT, CAS}
}
\email{zyq@ict.ac.cn}

\author{Kun Li}
\affiliation{%
  \institution{SKL of Computer Architecture, ICT, CAS}
}

\author{Pengqi Lu}
\affiliation{%
  \institution{SKL of Computer Architecture, ICT, CAS}
}

\author{Yue Yue}
\affiliation{%
  \institution{SKL of Computer Architecture, ICT, CAS}
}




\begin{abstract}
Stencil computations represent a very common class of nested loops 
in scientific and engineering applications.
Exploiting vector units in modern CPUs is crucial to achieving peak performance.
Previous vectorization approaches often consider the data space,
in particular the innermost unit-strided loop.
It leads to the well-known data alignment conflict problem 
that vector loads are overlapped due to the data sharing between continuous 
stencil computations.
This paper proposes a novel temporal vectorization scheme for stencils.
It vectorizes the stencil computation in the iteration space and assembles
points with different time coordinates in one vector.
The temporal vectorization leads to a small fixed number of vector reorganizations
that is irrelevant to the vector length, stencil order, and dimension.
Furthermore, it is also applicable to Gauss-Seidel stencils,
whose vectorization is not well-studied.
The effectiveness of the temporal vectorization is demonstrated
by various Jacobi and Gauss-Seidel stencils.
\end{abstract}


\maketitle

\section{Introduction}

The stencil computation is identified as one of the thirteen Berkeley motifs and
 represents a very common class
of nested loops in scientific and engineering applications, dynamic programming, and image processing algorithms.
A stencil is a pre-defined pattern of neighbor points used for updating a given point. 
The stencil computation involves time-iterated updates on a regular $d$-dimensional grid, 
called the \emph{data space} or \emph{spatial space}. 
The data space is updated along the time dimension, 
generating a $(d+1)$-dimensional space referred to as the \emph{iteration space}. 
The stencils can be classified from various perspectives, 
such as the grid dimensions (1D, 2D, ...), 
orders (number of neighbors, 3-point, 5-point, ...), shapes (box, star, ...), 
dependence types (Gauss-Seidel, Jacobi) and boundary conditions (constant, periodic, ...).

The naive implementation for a $d$-dimensional stencil 
is comprised of $(d+1)$ loops where the outermost loop traverses the time dimension
and the inner loops update all grid points in the $d$-dimensional spatial space.
It exhibits poor data reuse and is a typical bandwidth-bound kernel.
For improving the performance, blocking and vectorization
are the two most powerful and commonly used transformation techniques.

There are two kinds of blocking methods for stencil computations:
\emph{spatial blocking} and \emph{temporal blocking}. 
The spatial blocking algorithms promote data reuse in a single time step for 2D and higher dimension stencils
by adjusting the data traversal pattern to an optimized order.
An in-cache grid point may be reused to update all its neighbors before evicted from the cache.
However, the locality exploited by space blocking is limited by the neighbor pattern size of a stencil.
The temporal tiling \cite{Ding.He:sc01,Rastello.Dauxois:ipdps02,Rivera.Tseng:sc00,Nguyen+:sc10,Meng.Skadron:ics09,Yuan+:sc17}
takes the time dimension
into consideration simultaneously with spatial dimensions.
It has been exhaustively studied for stencils to further improve the data locality
and alleviate the memory bandwidth demands.

The vectorization groups a set of data in a vector register and processes them in parallel
to utilize vector units in modern CPU architectures.
It exploits the data parallelism and serves to boost the in-core performance.
There has been a long history of efforts to design efficient vectorization methods
\cite{Allen.Kennedy:toplas87,Allen.Kennedy:book01,Sreraman.Govindarajan:ijpp00,Larsen.Amarasinghe:pldi10,Hampton.Asanovic:cgo08, Nuzman.Zaks:pact08,Zhou.Xue:taco16,Maleki+:pact11}.
Though the stencil computation is characterized by its apparently low arithmetic intensity,
the vectorization is still profitable, especially for blocked stencil algorithms.
Prior vectorization techniques for stencils focus on the data space,
i.e. group points either in the uni-stride space dimension \cite{Henretty+:cc11} or
multiple space dimensions \cite{Yount:hpcc15}. 
We refer to this scheme as \emph{spatial vectorization}.

One well-known problem induced by the spatial vectorization of stencils
is the data alignment conflict. 
It arises from the fact that continuous vectors require
the same value appears at different positions of vectors.
Thus it incurs either redundant loads or additional data organization operations.
Existing solutions \cite{Caballero+:ics15,Henretty+:cc11}
reduce these overheads but 
still limit the performance or hurt the data locality.
We will provide a detailed analysis in Section \ref{section-background}.

Furthermore, vectorization and blocking are often regarded as two orthogonal methods.
However, the vectorization and tiling actually interact with each other for stencil computations.
The vectorization often requires higher bandwidth
and favors loading data from the first-level cache.
On the contrary, the tiling tries to minimize the data transfer at the memory-cache level and prefers the last-level cache.
This performance gap motivates our work and will be discussed in Section \ref{section-motivation}.


In this paper, we propose a novel \emph{temporal vectorization} scheme
considering the entire iteration space.
A vector in the temporal vectorization scheme groups points with different time coordinates. 
It seeks to alleviate the data alignment conflict
and bridges the above-mentioned performance gap.  
We also design a set of optimizations to 
alleviate weaknesses introduced by the new scheme
and adjust the data layout to explore its potential.

The temporal vectorization still requires data reorganization operations,
but the overhead is fixed and smaller than previous methods.
Furthermore, a vectorization scheme usually only affects the single-core performance.
However, the proposed temporal vectorization method leads to better utilization of the memory bandwidth. 
And in particular, it loads
the data at a slower speed.
Thus it can expect less memory contention especially for multi-core executions.
We implemented the temporal vectorization with temporal blocking schemes
and shows that the speedup increases with the number of cores especially 
for high-dimensional stencils.
Finally, the temporal vectorization scheme is also applicable
to the Gauss-Seidel stencils.
Gauss-Seidel stencils update a point using the newest values of neighbor points.
It is illegal to vectorize any single loop of the naive implementation code.
To the best of our knowledge, we are not aware of vectorization methods
for Gauss-Seidel stencils.

This paper makes the following contributions:

\begin{itemize}

\item We clarify a key characterization of stencils: the vectorization of stencils
    is sensitive to cache bandwidth and the blocking often favors a cache level with large capacity.
   The data alignment conflicts induced by vectorization enlarges this gap 
   and is still unsolved by existing methods.

\item We propose a novel temporal vectorization scheme for stencils.
It expands the target scope of vectorization from the spatial space to iteration space. 
The temporal vectorization incurs a smaller fixed number of data reorganization operations 
and leads to a better data transfer rate than previous methods.
Therefore it is less sensitive to the cache bandwidth. 
Furthermore, it is also applicable to Gauss-Seidel stencils.

\item The effectiveness of the temporal vectorization is demonstrated
by various Jacobi and Gauss-Seidel stencils.

\end{itemize}


The remainder of this paper is organized as follows.
Section \ref{section-background} provides the background.
The temporal vectorization is
described in Section \ref{section-algorithm}.
We present the performance results in Section \ref{section-result}.
Section \ref{section-relatedwork} overviews related work
and 
Section \ref{section-conclusion} concludes the paper.

\section{Background}
\label{section-background}

\subsection{Data Alignment Conflict of Vectorization}

\begin{algorithm}[t]
  \For{$t = 0;\ t < T;\ t=t+1$}{
  \For{$x = 1;\ x <= NX;\ x=x+1$}{
    $a_x^{t+1}$ = Stencil$(a_{x-1}^t,\,a_{x}^t,\,a_{x+1}^t)$\;}}
  \caption{1D3P Jacobi Stencil, scalar code}
  \label{alg-1}
\end{algorithm}

We take the 1D3P stencil as an example to illustrate the fundamental problem of the stencil 
computations caused by vectorization. 
The pseudo-code is listed in 
Algorithm \ref{alg-1}.
$a_{x}^{t}$ repersents the value at the point $(t,x)$ in the iteration space
where $x$ and $t$ are the coordinates in the space and time dimension, respectively.
In each iteration of the inner space loop, it loads $a_{x+1}^{t}$, 
reuses in-register data $a_{x-1}^{t}$ and $a_{x}^{t}$ referenced by the previous
 calculation and writes the result $a_{x}^{t+1}$ to memory. 
 Observing the CPU-memory data transfer, one iteration of the inner loop is 
 exactly similar to a common array copy algorithm.

The vectorization groups a set of data in a vector register and processes them in parallel.
The naive vectorization of the 1D3P stencil code computes contiguous elements in the output array 
$a_x^{t+1}$. Assume the vector register holds 4 elements (i.e. vector length $vl=4$), 
the vectorization code 
performs the calculation with vector operations and outputs 
$(a_1^{t+1},a_2^{t+1},a_3^{t+1},a_4^{t+1})$ using one 
vector register.

A well-known problem incurred by the vectorization of stencil codes is the input data alignment
 conflicts. For example, to compute $(a_1^{t+1},a_2^{t+1},a_3^{t+1},a_4^{t+1})$, it requires three vectors: 
 $(a_0^{t},a_1^{t},a_2^{t},a_3^{t})$, $(a_1^{t},a_2^{t},a_3^{t},a_4^{t})$ 
 and $(a_2^{t},a_3^{t},a_4^{t},a_5^{t})$. 
 The element $a_2^{t}$ appears in all these vector registers but at different positions. 


The fundamental reason for the data alignment conflict is the data sharing between continuous calculations,
e.g.,  $a_x^1$ and $a_{x+1}^1$ depend on same points  $a_x^0$ and $a_{x+1}^0$.
We refer to this as the \emph{read-read dependence}.
Conventionally the read-read dependence is not a data hazard as other data dependencies
including read-after-write, write-after-read, and write-after-write dependencies.
Furthermore, common read-read dependence is usually exploited to promote data locality.
However, for vectorized stencil codes, the data alignment conflict arises from the fact that components in one output vector
or at the different positions of some output vectors have intra-vector or inter-vector read-read dependencies.
Then the required data must redundantly appear in many vectors.

\subsection{Existing methods}
We present three existing solutions to the data alignment problem
and discuss their drawbacks.

\begin{algorithm}[t]
  \For{$t = 0;\ t < T;\ t=t+1$}{
  \For{$x = 1;\ x <= NX - 3;\ x=x+4$}{
      $v_0$ = vload$(a_{x-1}^t, a_{x}^t, a_{x+1}^t, a_{x+2}^t)$\;
      $v_1$ = vload$(a_x^t, a_{x+1}^t, a_{x+2}^t, a_{x+3}^t)$\;
      $v_2$ = vload$(a_{x+1}^t, a_{x+2}^t, a_{x+3}^t, a_{x+4}^t)$\;
    $(a_{x}^{t+1}, a_{x+1}^{t+1}, a_{x+2}^{t+1}, a_{x+3}^{t+1})$= Stencil($v_0,\,v_1,\,v_2)$\;
  }
  }
  \caption{1D3P Jacobi Stencil, multi-load code}
  \label{alg-2}
\end{algorithm}

\emph{Multiple load vectorization}.
The common vectorization employed by production compilers loads all the needed vectors from memory straightforwardly
as shown in Algorithm \ref{alg-2}. 
Due to the low operational intensity, the stencil computation is often regarded as a 
memory-starving application. Compared with the scalar code, this multiple load 
vectorization method further increases the data transfer volume. Moreover, 
in each iteration of this code, it has at least two unaligned memory references
where the first data address is not at a 32-byte boundary. Since CPU implementations 
favor aligned data loads and stores, these unaligned memory references will degrade the performance considerably.

\emph{Data reorganization vectorization}.
Another solution \cite{Caballero+:ics15,Zhou.Xue:cgo16} is similar to the scalar code in terms of the CPU-memory data transfer. 
It loads each input element to vector register only once and assembles the required vectors 
via inter-register data permutations instructions. Compared with the multiple load method, 
this data permutations method reduces the memory bandwidth usage and takes the advantages of 
the rich set of data-reordering instructions supported by most SIMD architectures. However, 
the execution unit for data permutations inside the CPU may become the bottleneck.

A common disadvantage of these two approaches is that the number of redundant data loads 
or reorganization operations increases with the order of a stencil, the length of the CPU vector register
and the dimensionality of the problem.
For example, to compute the vector $(a_1^1, a_2^1, a_3^1, a_4^1)$ of the 1D5P stencil,
it needs to put $a_3^0$ in four vectors at all different positions to update $a_1^1$, $a_2^1$, $a_3^1$ and $a_4^1$.
Thus the redundancy is proportional to the order of a stencil and at most $vl - 1$.
For the 2D9P stencil, the innermost loop incurs two redundant loads and the outer space loop
incur another four.

\emph{Dimension-Lifting Transpose} (DLT).
One milestone approach to address the data alignment conflict is the DLT method \cite{Henretty+:cc11}. 
It turns to put the points with read-read dependencies in the same position of different vectors.
Specifically, the original one-dimensional array of length $N$ is viewed as a matrix of size $vl*(N/vl)$. 
It then performs a matrix transpose. 
Consider the DLT method for a one-dimensional array of 28 elements. 
The second $vl=4$ elements in the transformed layout are contiguous stored and loaded into one output vector 
$(a_1^1, a_8^1, a_{15}^1, a_{22}^1)$. 
All the three required input vectors:  $(a_0^0, a_7^0, a_{14}^0, a_{21}^0)$,
$(a_1^0, a_8^0, a_{15}^0, a_{22}^0)$ and $(a_2^0, a_9^0, a_{16}^0, a_{23}^0)$
are free of data sharing 
and also stored contiguously in memory. DLT only needs to assemble input vectors for calculating output vectors at boundary. 


DLT has the following disadvantages. First, DLT can be viewed as $vl$ independent 
stencils if we ignore the boundary processing. Therefore when incorporated with
 blocking frameworks, the data reuse decreases $vl$ times. The reason is that there 
 is no data reuse among the $vl$ independent stencils. Second, DLT suffers from the 
 overhead of explicit transpose operations executed before and after the stencil computation. 
 For 1D and 2D stencils in scientific applications, the number of time loops is often large 
 enough to amortize the transpose overhead.
But for 3D stencils and low-dimensional stencils in other applications like image processing,
 the time size is often too small to amortize the overhead.
Third, it's hard to implement the DLT transpose in-place and it often chooses to use 
an additional array to store the transposed data. This increases the space complexity of the code.
Finally, DLT fails to apply to  Gauss-Seidel stencils.

\section{Temporal Vectorization}
\label{section-algorithm}

\subsection{Motivation}
 \label{section-motivation}

 Our work is motivated by two observations
 on the data transfers between the CPU and cache,
 and between the cache and memory.
 
 First, the vectorized codes often achieve higher performance than
 scalar codes. 
 Furthermore, the data alignment conflict induced by vectorization
 requires either more data transfers or data reorganization operations.
 Consequently, the vectorization increases the bandwidth demands.
 As will be demonstrated in the Evaluation section,
 all sequential non-blocking stencil implementations with existing vectorization techniques 
 achieve the highest performance when the
 problem sizes fit in the L1 cache
 and the performance decreases fast as the problem size increases.
 Since the L2 cache provides competitive bandwidth compared with the L1 cache,
 it implies that existing vectorization schemes are relatively cache-bandwidth-sensitive.
 
 Second, L1 cache sizes in modern CPUs are often small.
 The typical size is around 32 KB.
 It holds up to 4000 elements for a double-precision floating-point kernel.
 Thus for higher dimension stencils the space blocking sizes
 and the corresponding temporal blocking size are limited,
 which will lead to a high memory transfer volume.
 As will be demonstrated in the Evaluation section,
 the parallel blocking stencil implementations with existing vectorization techniques 
 often get the best performance when the block fits in L1 cache or L2 cache for
 one-dimensional or high-dimensional stencils.
 This demonstrates the memory-bandwidth-bound restriction should be first satisfied
 even it incurs a slower in-core performance for L2 cache.
 This indicates that the blocking scheme is cache-size-sensitive especially for high dimension stencils.

 These two observations illustrate the trade-off
 between the data locality exploitation at cache-memory level
 and the in-core performance of the vectorized codes
 at the CPU-cache level.
 The conventional innermost loop vectorization
 leads to the best data reuse at the memory-cache level
 while incurs redundant CPU-cache data transfers.
 
 The DLT generates an optimal in-core data access pattern 
 while hurts the data locality in the cache.
 The sequential DLT results \cite{Henretty+:cc11} exhibit performance improvements for all stencils when the data fit in caches.
 However, the DLT with a blocking scheme \cite{Henretty+:ics13} derives considerable speedups for 1D stencils
 but only competitive or even worse performance for high-dimensional stencils.
 This implies that the degradation of data locality overweights the benefits of the vectorization
 for the DLT.
 
 We seek to design a vectorization scheme that maintains the data reuse
 ability and incurs a lightweight in-core data reorganizations
 simultaneously.

\begin{algorithm}[b]
  \For{$t = 0;\ t < T;\ t=t+4$}{
      Compute($a^{t+1}_{1},\dots,a^{t+1}_{2+2s}$)\;
      Compute($a^{t+2}_{1},\dots,a^{t+2}_{2+s}$)\;
      Compute($a^{t+3}_{1},\dots,a^{t+3}_{2}$)\;
      $v_0$ = vloadset$(a_0^{t+3}, a_s^{t+2}, a_{2s}^{t+1}, a_{3s}^t)$\;
      $v_1$ = vloadset$(a_1^{t+3}, a_{1+s}^{t+2}, a_{1+2s}^{t+1}, a_{1+3s}^t)$\;
      $v_2$ = vloadset$(a_{2}^{t+3}, a_{2+s}^{t+2}, a_{2+2s}^{t+1}, a_{2+3s}^t)$\;
  \For{$x = 1;\ x <= NX + 1 - 4s;\ x=x+1$}{
    $v_3$ = Stencil($v_0,\,v_1,\,v_2)$\tcp*{$v_3=(a_{x}^{t+4}, a_{x+s}^{t+3}, a_{x+2s}^{t+2}, a_{x+3s}^{t+1})$}
    $v_0$ = $v_1$\;
    $v_1$ = $v_2$\;
    store($a_{x}^{t+4}$)\tcp*{Line 12-14 may be in different orders}
    $v_3$ = vrotate($v_3$)\tcp*{$v_3=(a_{x+s}^{t+3}, a_{x+2s}^{t+2}, a_{x+3s}^{t+1}, a_{x}^{t+4})$}
   $v_2$ = vblend($v_3$,$a_{x+4s}^{t}$)\tcp*{$v_2=(a_{x+s}^{t+3}, a_{x+2s}^{t+2}, a_{x+3s}^{t+1},  a_{x+4s}^{t})$}
  }
  store($v_0$)\tcp*{$v_0=(a_{NX-3s-1}^{t+3}, a_{NX-2s-1}^{t+2}, a_{NX-s-1}^{t+1}, a_{NX-1}^t)$}
  store($v_1$)\tcp*{$v_1=(a_{NX-3s+0}^{t+3}, a_{NX-2s+0}^{t+2}, a_{NX-s+0}^{t+1}, a_{NX+0}^t)$}
  store($v_2$)\tcp*{$v_2=(a_{NX-3s+1}^{t+3}, a_{NX-2s+1}^{t+2}, a_{NX-s+1}^{t+1}, a_{NX+1}^t)$}
      Compute($a^{t+1}_{NX}$)\;
      Compute($a^{t+2}_{NX-s},\dots,a^{t+2}_{NX}$)\;
      Compute($a^{t+3}_{NX-2s},\dots,a^{t+3}_{NX}$)\;
      Compute($a^{t+4}_{NX-3s},\dots,a^{t+4}_{NX}$)\;
  }
    \caption{1D3P Jacobi Stencil, temporal vectorization code with $s=2$}
    \label{alg-temporalvectorization}
\end{algorithm}

\begin{figure}[t]
  \centering
   \includegraphics[width=0.93\linewidth]{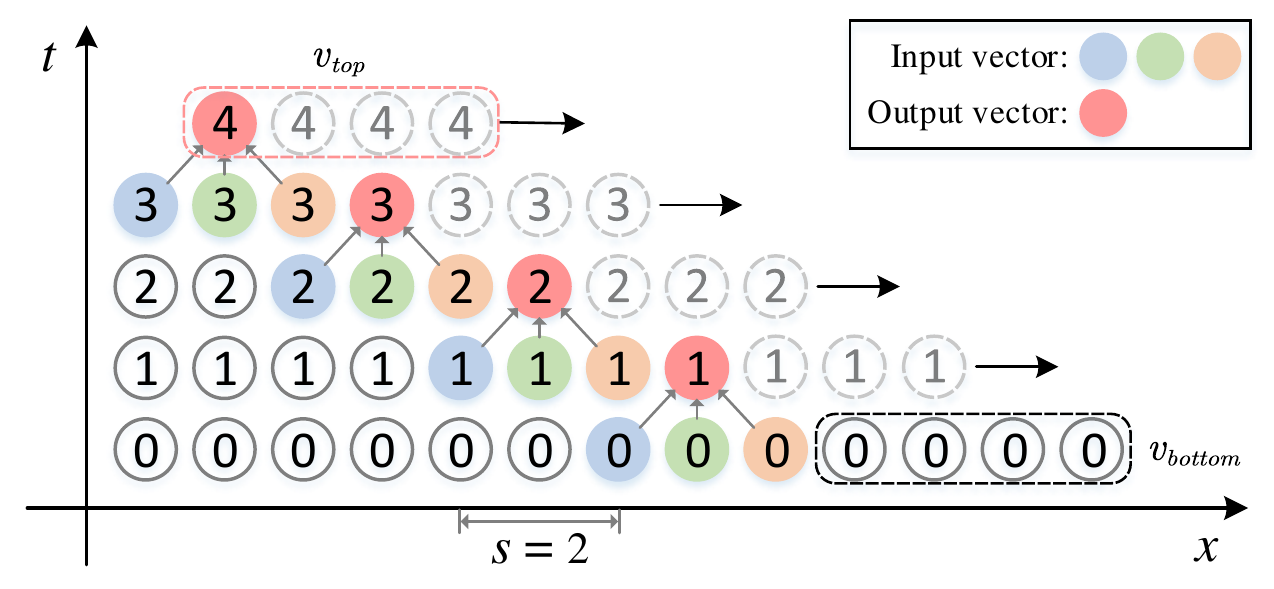}
  \caption{Temporal Vectorization of 1D3P Jacobi Stencil} 
\label{fig-temporalvectorization}
\end{figure}

\subsection{Algorithm}

The key idea is to extend the vectorization scope
from the data space to iteration space.
Specifically, data points with different time coordinates are assembled
in one vector.
We take the one-dimensional Jacobi of double-precision floating-point data as an example and assume one vector holds
four elements.
The general form of a vector encapsulating point values in an one-dimensional stencil is
$(a_{x_3}^{t_3}, a_{x_2}^{t_2}, a_{x_1}^{t_1}, a_{x_0}^{t_0})$.
For the stencil kernels used in the temporal vectorization, we always set $t_{i+1} - t_{i} = 1$.
The \emph{space stride} $s=x_{i} - x_{i+1}$ is determined by the stencil. 
Note that if it becomes the common vectorization in the multi-load and data reorganization
approaches when $t_{i+1}=t_{i}$ and $s=1$, and DLT when $t_{i+1}=t_{i}$ and $s = NX/4$.

To perform the stencil computation with a vector it only requires the elements in the vector
are free of dependence.
This is equivalent to respect the dependence defined by a stencil.
Let the dependence set of a stencil be $D$.
 Each dependence $(dt, dx)\in D$ implies that
there exists two points $(t', x')$ and $(t'',x'')$
with $t''-t'=dt$ and $x''-x'=dx$ 
in the iteration space that  $a_{x''}^{t''}$ depends
on $a_{x'}^{t'}$.
It is easy to show that the temporal vectorization 
is legal when $s>\max \{dx/dt | (dt, dx)\in D, dx >0\}$.
We only consider dependencies with $dx>0$
since the innermost loop traverses the space dimension
in increasing order.
Take the 1D3P Jacobi stencil as an example,
the dependencies are $(1, 0)$, $(1, 1)$ and $(1, -1)$.
Thus it is sufficient to make the temporal vectorization legal
by setting $s>1$.

Algorithm \ref{alg-temporalvectorization}
shows the pseudo-code of the temporal vectorization
of the 1D3P Jacobi stencil with the space stride $s=2$.
Lines 2-4 update a small set of grid points at time $t+1$,
$t+2$ and $t+3$
since the temporal vectorization
assembles some of these pre-computed values of different time and space coordinates.
Lines 5-7 collect vectors for the first vectorized stencil computation.
These vectors are called \emph{input vectors}
since they are fed to the vectorized stencil computing.
Note that the values in one input vector are not stored contiguously in memory.
So it must use strided load operations (e,g, gather or \verb=_mm256_set_pd= intrinsics in Intel AVX).
Line 9 performs the calculations and generates an
\emph{output vector}.

The temporal vectorization only requires the data reorganization of
the \emph{output vector} $(a_{x}^{t+4}, a_{x+s}^{t+3}, a_{x+2s}^{t+2}, a_{x+3s}^{t+1})$.
Since the next iteration of the outer time loop only requires the values
with time coordinate $t+4$,
the component $a_{x}^{t+4}$ at the highest position of the output vector
is the actual output value of the innermost loop and must be stored to memory (Line 12). 
The other components should be moved to their left (higher) positions (Line 13) and blended with
a new input element $a_{x+4s}^{0}$ (Line 14). The assembled \emph{input vector} 
$(a_{x+s}^{3}, a_{x+2s}^{2}, a_{x+3s}^{1},a_{x+4s}^{0})$
is used in subsequent stencil computations.
Note that this vector can be either preserved in a CPU vector register or output to the memory
for further use.
Finally, Lines 19-22 calculate the rest values that are not covered by the vectorization code.
Figure \ref{fig-temporalvectorization} illustrates the Algorithm.
Note that the points with a same color are assembled in one vector.

This algorithm iteratively sweeps a time tile of the height equivalent to the vector length 4
and forwards all grid points from time coordinate $t$ to $t+4$ in one iteration of the inner loop
at Line 8.
All the points in the iteration space with time coordinate $t+i$ ($t=4k, i=0,1,2,3$)
always appears at the $i$-th position of one input vector.
Thus in the rest of the paper
we can ignore $t$ and always use $(a_{x}^{3}, a_{x+s}^{2}, a_{x+2s}^{1}, a_{x+3s}^{0})$
represents an input vector and 
$(a_{x}^{4}, a_{x+s}^{3}, a_{x+2s}^{2}, a_{x+3s}^{1})$
an output vector.

The values at the highest position of the output vectors in every four continuous iterations of the innermost loop
are assembled in one \emph{top vector} $v_{top}=(a_{x+3}^4, a_{x+2}^4, a_{x+1}^4, a_x^4)$ 
and written to memory with a vector-storing instruction
as shown in Figure \ref{fig-temporalvectorization}.
This task needs 5 data reorganization instructions.
The first two values $a_x^4$ and $a_{x+1}^4$
are copied to the two lower positions of $v_{top}$ after rotating their corresponding output vectors. 
The last two output values $a_{x+2}^4$ and $a_{x+3}^4$
are copied to the two higher positions of $v_{top}$ before rotating their corresponding output vectors.

Similarly, the \emph{bottom vector} $v_{bottom}=(a_{x+4s+3}^0$, $a_{x+4s+2}^0$, $a_{x+4s+1}^0$, $a_{x+4s}^0)$
containing four continuous values with time coordinate 0
is loaded from memory by a 
vector-loading instruction and blended with four output vectors
calculated in four continuous iterations to generate four corresponding input vectors.
This task also requires 5 data reorganization operations.
The temporal vectorization also needs one data reorder instruction 
to rotate the output vector or input vector.
Thus the number of data reorganization operations per output vector is  $1+10/4=3.5$.

\begin{figure}[t]
  \centering
   \includegraphics[width=0.73\linewidth]{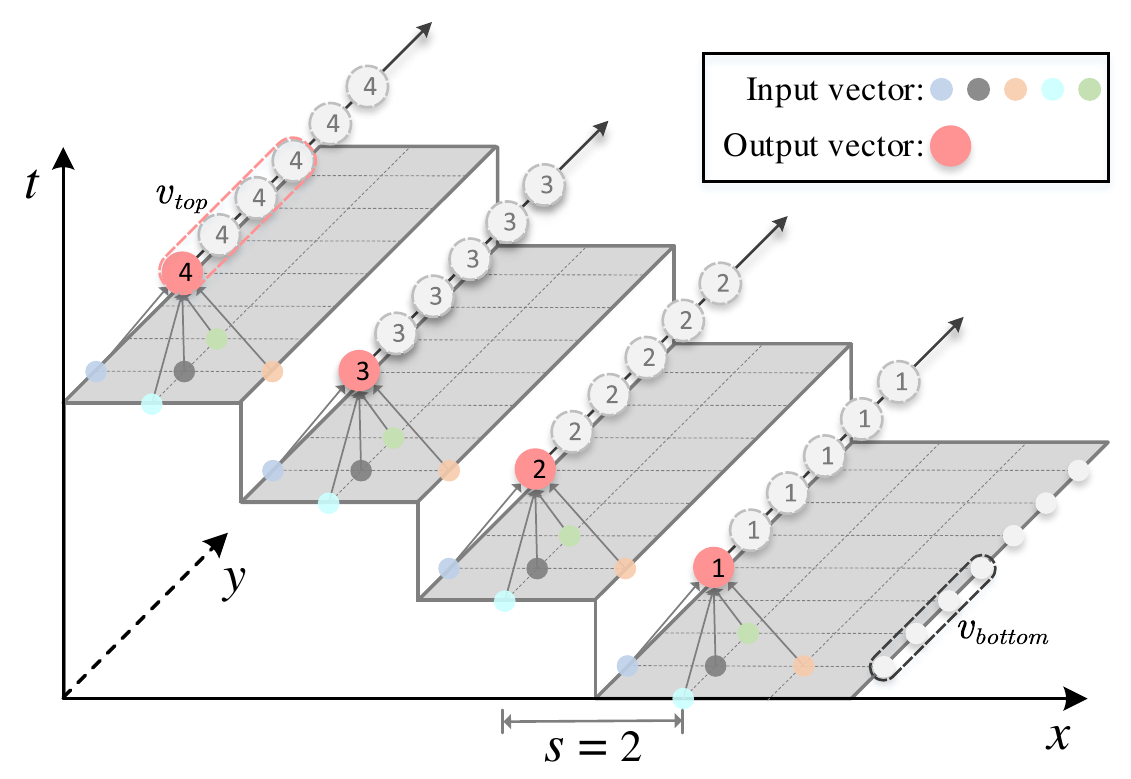}
  \caption{Temporal Vectorization of 2D5P Jacobi Stencil} 
\label{fig-temporalvectorization2}
\end{figure}

\emph{High-dimensional Stencils}.
The temporal vectorization for one-dimensional stencils
in effect performs a strip-mining to the outermost time loop
and turns it into two nested time loops.
The outer time loop index is incremented by the vector length
while the inner time loop traverses a time tile.
The temporal vectorization reorders the calculations 
in the inner time loop and the space loop.
For high-dimensional stencils, 
it is illegal to interchange the inner time loop
with space loops, thus it is only allowable to perform
the temporal vectorization on the outermost space loop.
Consequently, the pre-computation in Line 2-4 of Algorithm \ref{alg-temporalvectorization}
forwards grid points in several lines for 2D stencils or planes for 3D stencils
1, 2, or 3 time steps.
Figure \ref{fig-temporalvectorization2} shows a pictorial view of
the temporal vectorization of the 2D5P stencil.

The data transfer and vectorized computation in Line 5-18 of Algorithm \ref{alg-temporalvectorization}
can be easily extended to higher dimensions.
Note that there are additional space loops inside the $x$ loop in Line 8.
Therefore the reorganized input vectors, 
e.g. $(a_{x+s,y}^{3}, a_{x+2s,y}^{2}, a_{x+3s,y}^{1}, a_{x+4s,y}^{0})$ in a 2D stencil must be stored to memory for the next iteration of 
outer space loops, e.g. the computations in $x+s-1$, $x+s$ and $x+s+1$ iterations.

\subsection{Optimization}

We now present several drawbacks introduced by the temporal vectorization
and corresponding optimizations.


\emph{Efficient data reorganization}.
The first one is related to the data reorganization operations inside 
one iteration of the innermost loop in Algorithm \ref{alg-temporalvectorization}.
These operations must be executed sequentially. 
For example, to process the first output vector $(a_{x}^{4}$, $a_{x+s}^{3}$, $a_{x+2s}^{2}$, $a_{x+3s}^{1})$,
it needs permute it to $(a_{x+s}^{3}, a_{x+2s}^{2}, a_{x+3s}^{1}, a_{x}^{4})$,
copy out $a_x^4$ and copy in $a_{x+4s}^{0}$.
Note that the latter two operations have data dependence while the first reorganization can be 
arbitrarily permuted.
Nevertheless, these three instructions have to be executed in order.
Furthermore, 
in modern CPU architectures, the vector register is split to 
some 128-bit lanes. Data movements inside lanes
incur a lower latency, typical $1$ cycle versus $3$ for lane-crossing instructions.
The data reorganization instructions involving the bottom or top vectors
are in-lane operations that incur small latency.
Other instructions manipulating the output or input vectors solely,
e.g., the rotation operation in Line 13 of Algorithm \ref{alg-temporalvectorization},
are lane-crossing.
Therefore these three instructions lead to a total latency of 5 cycles.

To reduce the latency, our key observation is that only $a_{x+2s}^{2}$ in the output vector is moved across lanes
for the corresponding input vector assembling.
To reduce the latency of the data reorganization of one output vector
and the number of lane-crossing instructions,
we turn to copy out the value with the time coordinate $2$
and copy in a new one which is already in the high lane.

\begin{figure}[t]
  \centering
   \includegraphics[width=0.73\linewidth]{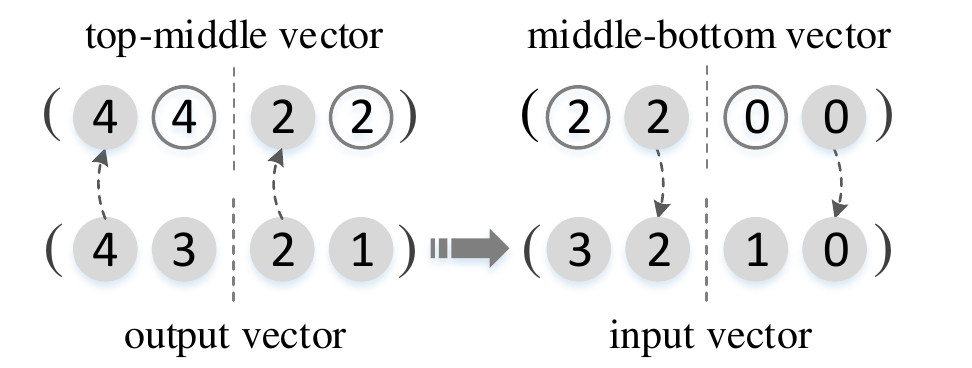}
  \caption{Reducing Lane-crossing Instructions} 
\label{fig-opt-lane}
\end{figure}

To achieve this, we add another space stride
(denoted as $sl$) between lanes of the vector.
Specifically, the form of output vectors becomes $(a_{x}^{4}, a_{x+s}^{3}$, $a_{x+2s + sl}^{2}, a_{x+3s+sl}^{1})$.
To form the corresponding input vector, both $a_{x}^{4}$ and $a_{x+2s + sl}^{2}$ are copied out
to a \emph{top-middle vector}.
Then it is blended with a \emph{middle-bottom} vector containing $a_{x+2s}^{2}$ and $a_{x+4s + sl}^{0}$ at different lanes
to form the corresponding input vector $(a_{x+s}^{3}, a_{x+2s}^{2}, a_{x+3s + sl}^{1}, a_{x+4s+sl}^{0})$.
Thus the number of data reorganization instructions on the critical path is decreased to 2
and both of them can be implemented with in-lane instructions.
Figure \ref{fig-opt-lane} illustrates this optimization.
In this paper, we always set $sl=2$.

To simplify the explanation of the top-middle and middle-bottom vector manipulations,
the space coordinates are ignored. It is easy to extend it with $x$ index and obtain the 
complete process. Every four iterations of the innermost loop generate two
top-middle vectors of the form $(a^4, a^4, a^2, a^2)$. They can be combined with
a bottom vector $(a^0,a^0,a^0,a^0)$ to produce two middle-bottom vectors
of the form $(a^2,a^2,a^0,a^0)$ for subsequent input vector assembling.
Finally, the two top-middle vectors are merged into a top vector.
It needs 3 lane-crossing instructions to reorganize the top-middle and middle-bottom vectors,
and the input and output vectors are no longer needed to be reorganized by lane-crossing instructions.
In sum, the total number of data reorganization instructions is reduced
to $3/4$ lane-crossing and 2 in-lane instructions per vectorized stencil computation.

\emph{Improving data parallelism for one-dimensional stencils}.
The temporal vectorization also introduces dependence between input and output vectors
in different iterations of the innermost loop.
The reason is that there is data dependence along the time dimension
and our vectorization scheme incorporates that dependence in iterations of the innermost space loop.
For example, after calculating the output vector
$(a_{1}^{4}, a_{3}^{3}, a_{5}^{2},a_{7}^{1})$ of the current iteration $x=1$,
the calculation of $a_{2}^{4}$ in the output vector 
$(a_{2}^{4}, a_{4}^{3}, a_{6}^{2},a_{8}^{1})$
of the next iteration $x=2$
depends on $a_{3}^{3}$.
This dependence resembles the common true (read-after-write) dependence.
It limits the number of concurrent instructions and may extremely impact performance.

One straightforward approach to increase the number of concurrent computations
for the temporal vectorization of one-dimensional stencils is 
to widen the space stride $s$.
As Figure \ref{fig-temporalvectorization} shows, the number of input vectors
for one-dimensional Jacobi stencils 
is $s+r$ where $r$ is half of the stencil order.
To determine the space stride $s$,
one consideration is about the data reorganization process.
Since a top vector groups the value of time coordinate $4$ in every four output vectors,
the number of available input vectors should be dividable by $4$
to simplify the algorithm implementation.
Another limitation to the upper bound of $s$ is the number of available vector registers in the CPU.
We conducted the experiments on CPUs with the AVX extension where 
the size of the vector register file is 16.
Take the top, bottom, top-middle, middle-bottom and coefficients vectors
into consideration, we always set the number of available input vectors
to 8.
Thus for the 1D3P Jacobi stencil, we set $s=7$.

\emph{Transposed data layout for high-dimensional stencils}.
The temporal vectorization of one-dimensional stencils is able to 
keep the input vectors in CPU vector registers.
Consequentially there is no memory transfer for the values with time coordinates 1, 2 and 3 computed
in the inner loop.
However, as explained above, 
since the temporal vectorization targets the outermost space loop of a high-dimensional stencil,
the input vectors must be stored to memory for subsequent stencil computations.
It is desirable to store the input vector contiguously.

Although the values in one input vector are not adjacent in the data space,
the values with the same time coordinate at the same position of these input vector are stored contiguously in memory.
For example, for four continuous input vectors
$(a_{x+s,y+i}^{3}$, $a_{x+2s,y+i}^{2}$, $a_{x+3s,y+i}^{1}$, $a_{x+4s,y+i}^{0})$,
 the four values with time coordinate 3, $a_{x+s,y+i}^{3}$ $(i=0,1,2,3)$
logically occupy continuous positions.
These four input vectors can be contiguously store in
these spaces and actually can be viewed as a transposed layout.

\emph{Initial input vectors loading (final output vectors storing)}.
The previous optimization improves the storage of the computed input vectors
for high-dimensional stencils.
The initial input vectors loading (Line 2-4 in Algorithm \ref{alg-temporalvectorization}) and final input vectors storing
(Line 16-18)
can be implemented in a similar approach.
These values are loaded to vectors by basic vector read instructions
(e.g. \verb=_mm256_load_pd=).
Each vector contains values with a same time coordinate.
Every four input vectors can be obtained by transposing corresponding vectors.
Similarly, the final input vector is stored in memory after a transpose.

\subsection{Implementation}

We now present implementations
of various stencils evaluated in this paper.
The combination of the temporal vectorization
and time-space blocks employed in parallelizations
are also described.

\emph{Jacobi stencils}.
We implemented 1D3P, 2D5P, and 3D7P Heat equation stencils
and a 2D9P stencil
with non-periodic boundary conditions.
They serve as the most important and most commonly used kernels in the study of 
compiler optimization, blocking scheme design, and vectorization of stencils.
Applying the temporal vectorization with the optimizations to these Jacobi stencils
is straightforward.
For the 1D3P stencil, we set $s=7$ and use 13 vector registers including
$7$ input vectors, one top-middle vector, one middle-bottom vector,
one top vector, one bottom vector, and two vectors for constant coefficients.
For the 2D5P, 2D9P and 3D7P stencil we set $s=2$.

For the parallel implementation,
we use the diamond tiling \cite{Bondhugula+:pldi08}
for the 1D3P stencil.
Extending the non-blocking implementation
to the diamond tiling code is simple.
Each time tile of height 4 of a diamond is a trapezoid
or inverted trapezoid. It only needs to change the 
loop boundary conditions for the blocking code implementation.
For high-dimensional stencils, we adopt a diamond and parallelogram \cite{Wonnacott:ijpp02} hybrid tiling.
The diamond tiling always applies to the outermost space loop
and co-works with the temporal vectorization.
The parallelogram tiling is performed on the rest space dimensions.
For the space dimensions with parallelogram tiling,
 the corresponding indices must be shifted in input and output vectors.
For example, the general form of the input vector becomes 
$(a_{x,y}^{3}, a_{x+s,y+1}^{2}, a_{x+2s,y+2}^{1}, a_{x+3s,y+3}^{0})$.
It is still straightforward to apply the optimizations.


\emph{Game of life} (Life).
Conway’s Game of Life is a 2D9P Jacobi stencil where the stencil computation 
depends on the values of 8 neighbors. 
We adopt a particular variant called B2S23 used in Pluto \cite{Bondhugula+:pldi08}.
Although it is sufficient to use one bit to represent the value (false for dead or true for live) at each grid point,
we use the integer type like other works to facilitate
the summation of values of 8 neighbors.
The implementation of temporal vectorization and
parallelization is similar to 2D5P and 2D9P Jacobi stencils.

\emph{Gauss-Seidel stencils}.
Gauss-Seidel stencils use the newest neighbor values
to update one point.
Conventionally it only requires one array to store
the lastest values of all grid points
 as opposed to two arrays in Jacobi stencils.
Another difference between Jacobi and Gauss-Seidel 
stencils is that the latter contains 
data dependencies in all time and space loops.
Thus it is illegal to perform the conventional innermost vectorization.
Nevertheless, the implementations of Gauss-Seidel stencils 
are similar to Jacobi stencils. 
For the neighbors whose newest values are used in the calculation,
the temporal vectorization uses their corresponding output vectors.
It also leads to the same data reorganization cost to Jacobi stencils.
We omit the detailed explanation.
It is also illegal to employ the diamond tiling for Gauss-Seidel stencils.
Thus we utilize parallelogram tiling for all space dimensions.

\emph{Longest common subsequence} (LCS).
LCS is a classic problem solved by the dynamic programming method.
It can be viewed as an 1D Gauss-Seidel stencil where
the value $lcs[x][y]$ on the point $(x,y)$ in the iteration space
represents the length of the longest common subsequence
of two sequences $A[1\dots x]$ and $B[1\dots y]$.
$lcs[x][y]$ depends on $lcs[x-1][y]$, $lcs[x-1][y-1]$, $lcs[x][y-1]$, $A[x]$ and $B[y]$.
Thus the space stride must satisfy $s\geqslant 1$.
To reduce the storage size, we view the $x$ loop as the time dimension and $y$ loop as space.
Consequently the sequence $B$ can be regarded as a variable coefficient. 
LCS allows the rectangle tiling in the iteration space. 
The implementation allocates two arrays $lcsA[x]$ and $lcsB[y]$
to store the values on the wavefront, i.e. the top and right boundaries of a rectangle.

\subsection{Discussion}

Compared with the multi-load vectorization method,
our temporal vectorization method incurs no redundant data transfer.
For one-dimensional stencils each value $a_x^t$ only appears in one input vector.
Furthermore, it is easy to align all the top and bottom vector transfers.
For high-dimensional stencils $a_{x,y}^{t}$ may be loaded
in serveral continuous iterations ($y-1$, $y$ and $y+1$ for the 2D5P stencil) of the next-to-innermost space loop,
but it still requires fewer data transfers to the multi-load vectorization.

Compared with the data reorganization approach, the number of data
reorganization operations in our method is fixed and irrelevant
to the vector length, the stencil order, and dimensionality.
With the double-stride optimization,
the temporal vectorization incurs 0.75 lane-crossing and
2 in-lane instructions.
The data reorganization vectorization needs 1 lane-crossing and
2 in-lane instructions for the 1D3P stencil and more
for higher-order and dimension stencils.

Compared with the DLT method, the temporal vectorization gives rise to a better
reuse of the data in cache.
Though the temporal vectorization also incorporates
data transpose operations, it only processes a small set of points 
and the overhead can be amortized by stencil calculations.

The temporal vectorization, data reorganization approach, and DLT methods achieve the same reuse pattern to the scalar code.
For example, it only needs to load one input vector to compute an output vector
for the 1D3P stencil. 
Thus the memory transfer volumes of their blocking implementations are also similar.
However, the temporal vectorization leads to slower transfer speed. Specifically, the straightforward vectorization
method touches all the input points in the data space in one time step
while the temporal vectorization loads these data in four time steps.
It can then expect less memory bandwidth contention, especially in multi-core executions.

Furthermore, for the two arrays in Jacobi stencils,
the output data $a_x^4$ and input data $a_x^0$ actually share the same array.
The input vectors can be stored a fixed range of the other array.
Thus it actually uses one array for non-blocking Jacobi implementations 
and only stores the data in two arrays at boundaries for blocking implementations.
Therefore the memory transfer volume is reduced by a factor of 2 for Jacobi stencils.

The temporal vectorization would be represented with a set of loop transformations,
i.e. the strip-mining of the time loop, 
the time skewing of the inner time loop and outermost space loop,
the loop-peel and finally the outer-loop vectorization.
However, there are some difficulties to implement it with general compiler techniques.
First, the temporal vectorization needs auxiliary variables, e.g. the extra arrays in
the static coefficient method or the $lcsA[x]$
in LCS. These auxiliary data often needs manual efforts \cite{Satish+:isca12}.
Second, it often requires a transposed data layout for efficient vector loads
for high-dimensional stencils that complicate the compiler transformations.
Third, for one-dimensional stencils the data with time coordinates 1, 2 and 3 are 
reused in CPU registers and not stored into memory.
Conventionally compilers are conservative to store data to memory as performed by the original code.
Finally, for high-dimensional stencils, it is illegal to interchange the time loop and spaces loops,
this complicates the outer-loop vectorization since it must be applied to a loop that contains more than
one inner loop.


Nevertheless, given the temporal vectorization algorithms,
it is straightforward to implement a code generator.
As a comparison, the DLT method can be viewed as a combination 
of the strip-mining of the innermost loop
and the outer-loop vectorization of the two innermost loops. 
A domain-specific framework for 
temporal vectorization of stencil computations can be designed
similarly.

The temporal vectorization also resembles the wavefront method \cite{Wolfe:ijpp86}
(loop skewing).
The wavefront method utilizes the parallelogram block shape in temporal tiling.
The temporal tiling aims at exploiting the data reuse in caches
and reducing the memory transfer volume,
while the temporal vectorization primarily serves to lessen the bandwidth pressure
between the CPU and cache.

\section{Evaluation}
\label{section-result}

	\begin{table}[b]
	  \centering
	\caption{Problem and Blocking Sizes}{
	\label{tab-benchmark}
  \resizebox{0.9\columnwidth}{!}{
	\begin{tabular}{|c|c|c|c|}
		\hline
         Benchmark & Problem Size & our blocking & auto blocking\\
   	\hline
Heat-1D &	$16000000\times6000$   &  $16384\times 128$ &  $2048\times 128$ \\
   	\hline
Heat-2D &	$8000^2 \times 2000$  &  $256^2 \times 64$ &  $256^2\times 64$\\
         \hline
2D9P &	$8000^2 \times 2000$  &  $256^2 \times 64$ &  $256^2\times 64$\\
         \hline
Heat-3D &	$800^3 \times 200$ &  $32^3 \times 8$ &  $32^3 \times 8$\\
   	\hline
Life &	$8000^2 \times 2000$&  $256^2 \times 32$ &  $128^2\times 32$\\
         \hline

GS-1D &	$16000000\times6000$   &  $2048\times 64$ &  $2048\times 64$ \\
   	\hline
GS-2D &	$8000^2 \times 2000$  &  $128^2\times 32$ &  $128^2\times 32$\\
         \hline
GS-3D &	$800^3 \times 200$ &  $32^3\times 32$ &  $32^3\times 32$ \\
         \hline
LCS &	$200000\times 200000$   &  $4096\times 4096$ &  $4096\times 4096$ \\
   	\hline
	\end{tabular}}
  }
	\end{table}

\subsection{Setup}

Our experiments were conducted on a machine made of two Intel Xeon E5-2670 processors with 2.70 GHz clock speed.
Each processor contains 12 physical cores and 
each core owns a 32KB private L1 cache, a 256KB private L2 cache, and a unified 30MB shared L3 cache.
We compiled the program with the ICC compiler version 19.1.1, using the optimization flag `-O3 -xHost'.

For each stencil kernel, we implemented a sequential code without any blocking scheme
and a parallel code with a specific blocking scheme described above.
The results of sequential codes exhibit the sensitivity to cache bandwidth.
and serve to determine the blocking sizes for the parallel experiments.
The paralle codes were scaled from uni-core to all the 24 cores. 
The problem sizes for various benchmarks are listed in Table \ref{tab-benchmark}. 
We simply tested all blocking sizes that are the power of two and symmetric for all spatial dimensions
and show the one producing the best performance in Table \ref{tab-benchmark}.
However, we observed that the performance is very sensitive to the tile sizes,
but this requires significant effort in auto-tuning and should be done separately.

We compared our scheme with ICC auto-vectorization.
The scalar performance is also plotted.
The performance is reported using Gstencils/s,
i.e. the number of points updated per second.


\subsection{Results}

\begin{figure}[b]
  \centering
   \subfloat[Heat-1D Sequential] {
   \includegraphics[width=0.47\linewidth]{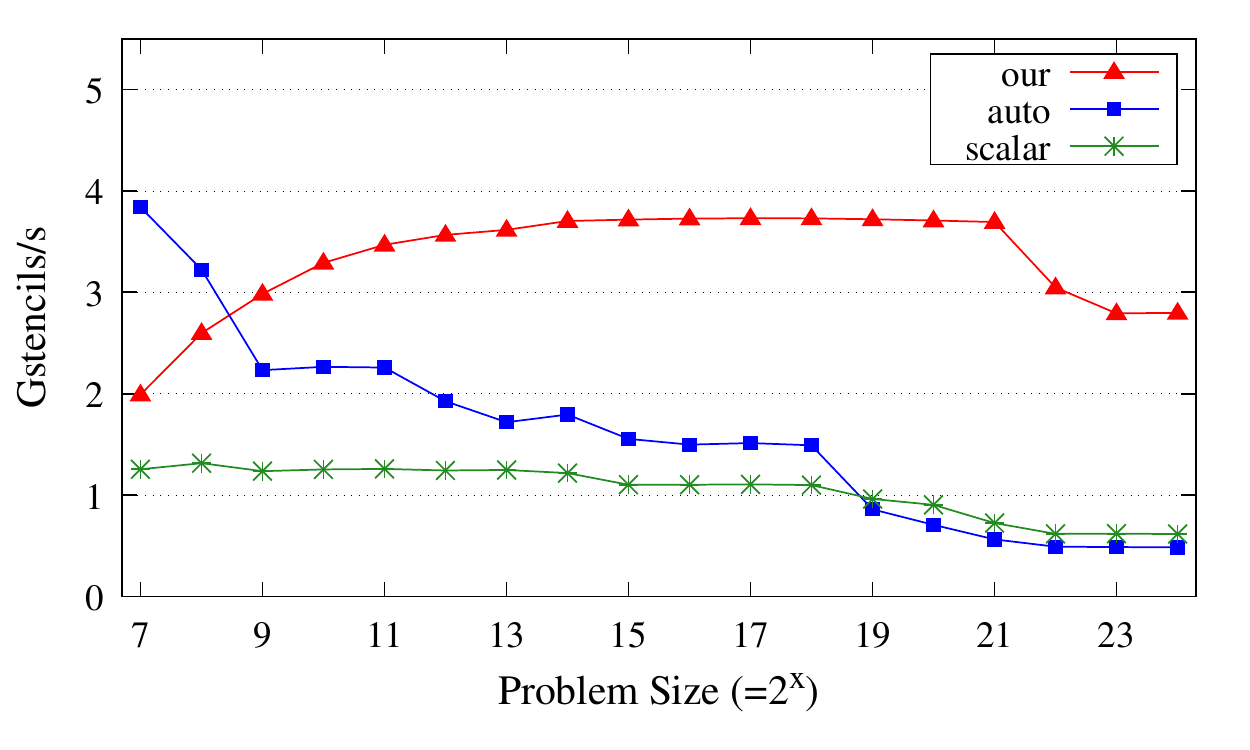}
   \label{fig-1d-jacobi-1}
  } 
\hfill
\subfloat[Heat-1D Parallel] {
   \includegraphics[width=0.47\linewidth]{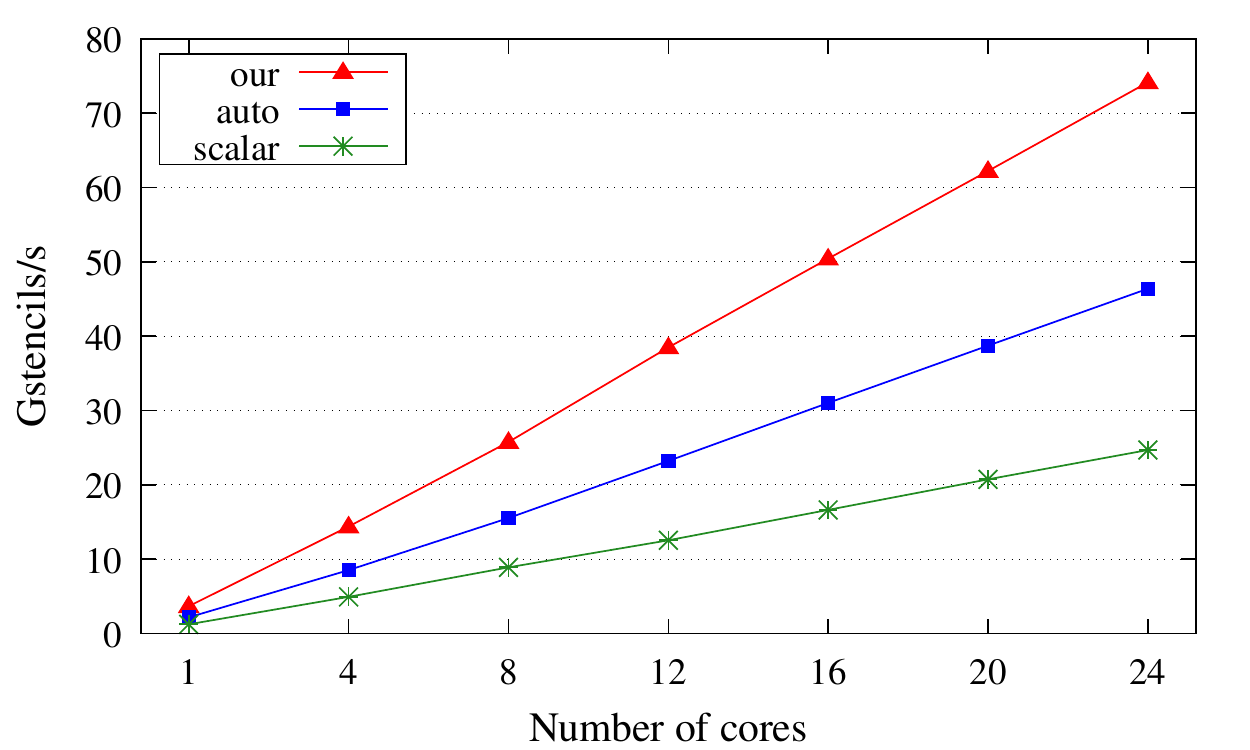}
   \label{fig-1d-jacobi-2}
  }
\hfill
 \subfloat[Heat-2D Sequential] {
   \includegraphics[width=0.47\linewidth]{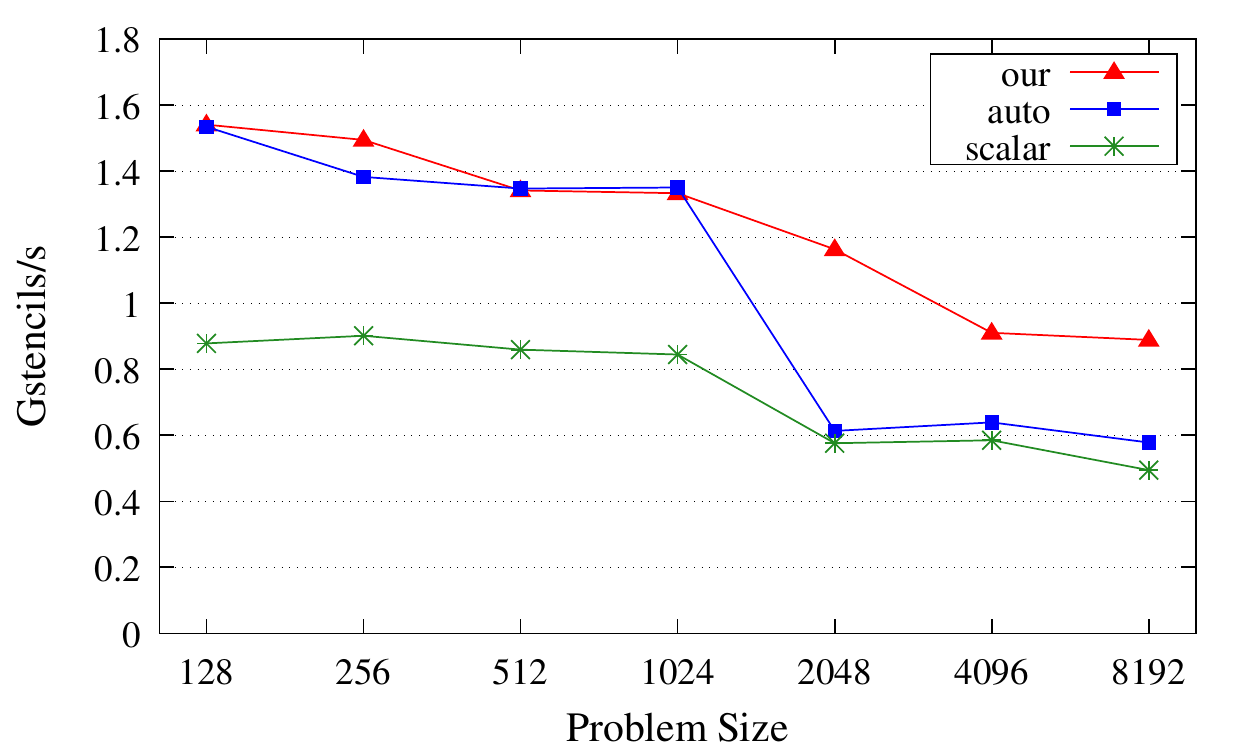}
   \label{fig-2d-jacobi-1}
  } 
\hfill
\subfloat[Heat-2D Parallel] {
   \includegraphics[width=0.47\linewidth]{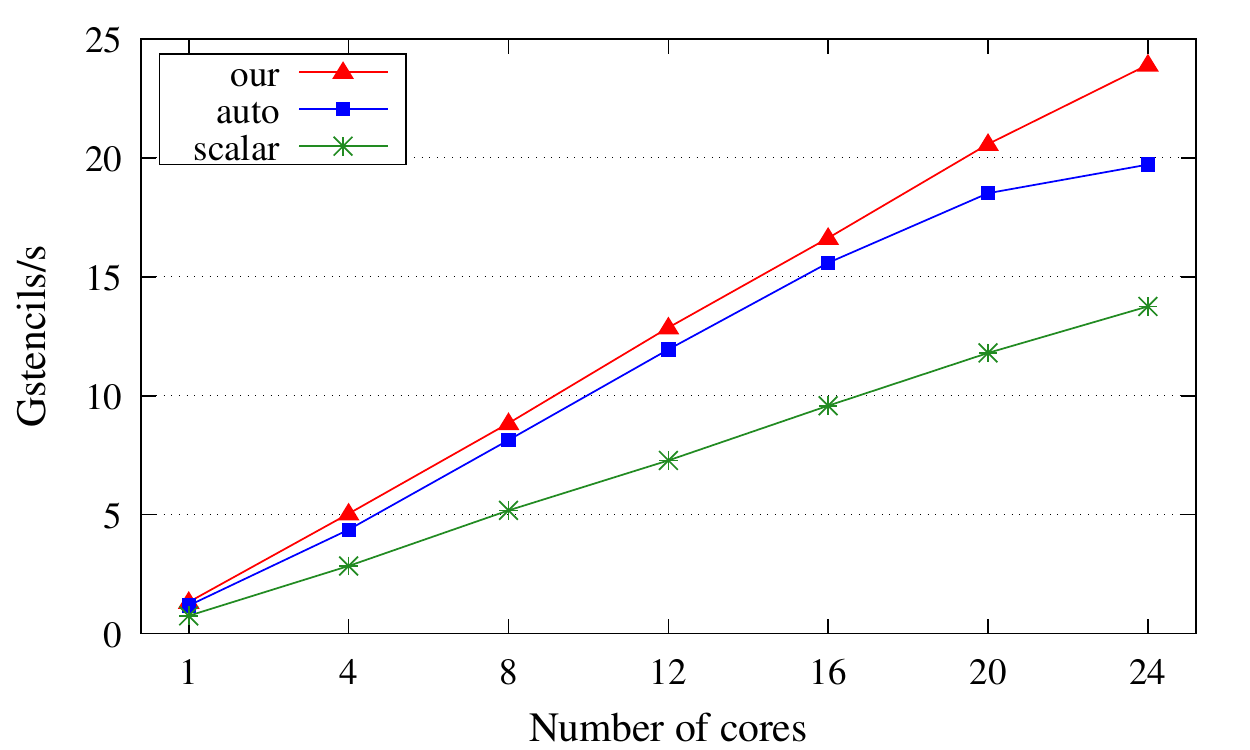}
   \label{fig-2d-jacobi-2}
  }
\hfill
 \subfloat[Heat-3D Sequential] {
   \includegraphics[width=0.47\linewidth]{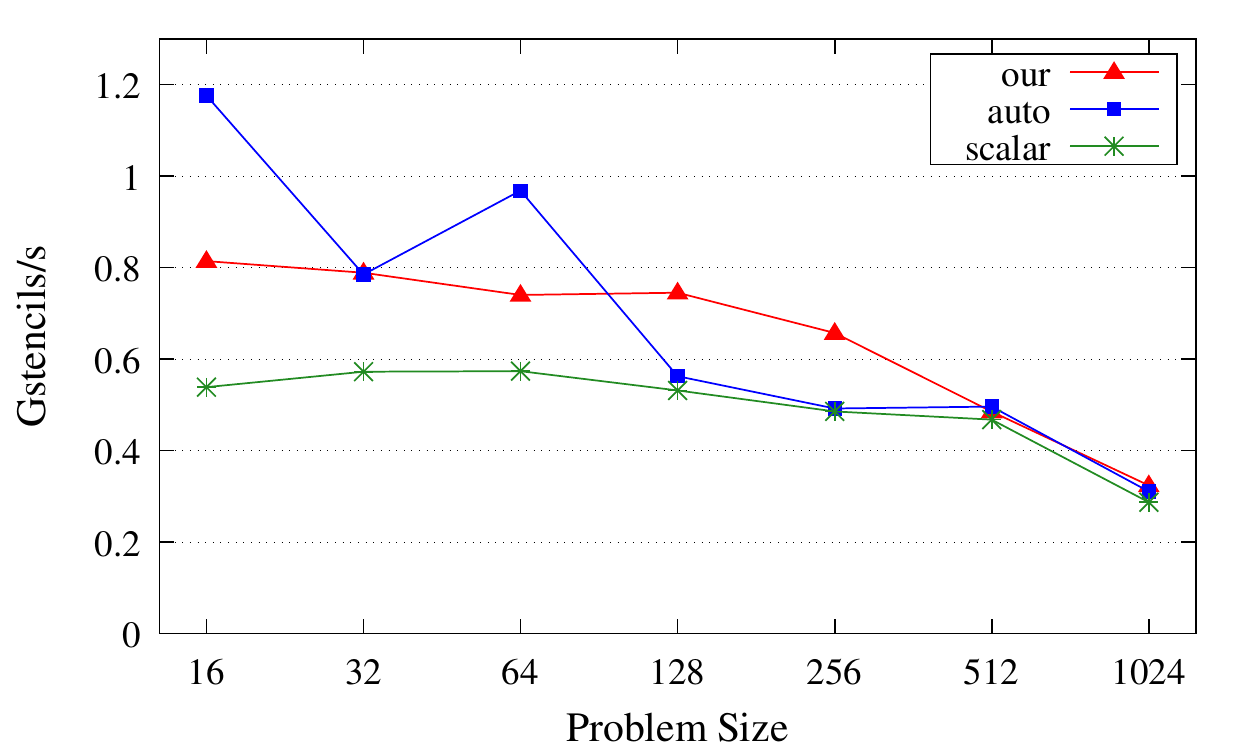}
   \label{fig-3d-jacobi-1}
  } 
\hfill
\subfloat[Heat-3D Parallel] {
   \includegraphics[width=0.47\linewidth]{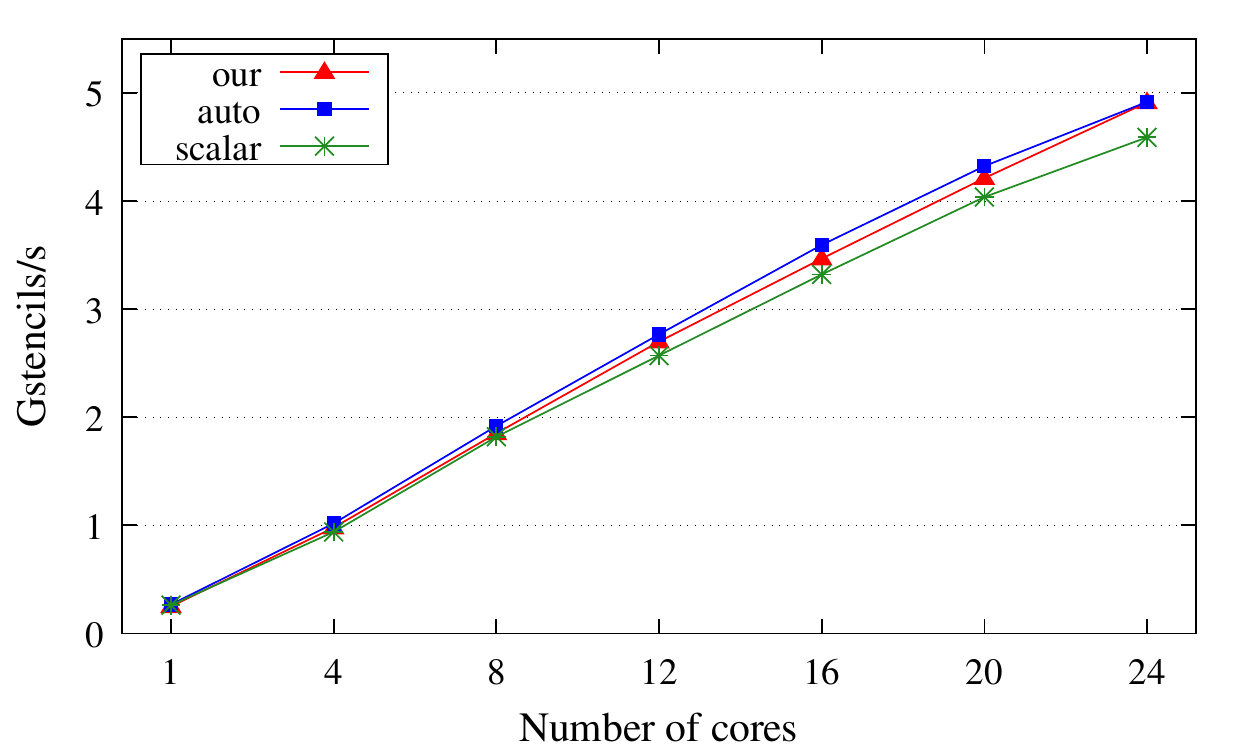}
   \label{fig-3d-jacobi-2}
  }
\hfill
 \subfloat[2D9P Sequential] {
   \includegraphics[width=0.47\linewidth]{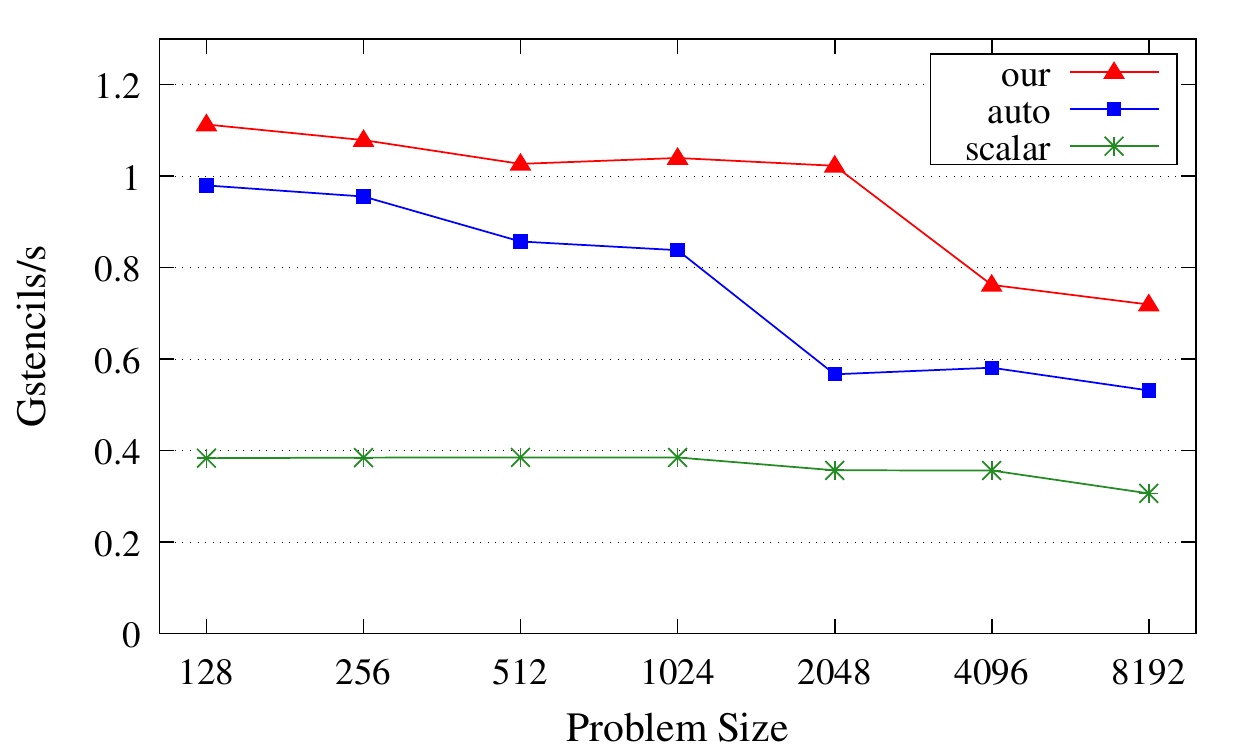}
   \label{fig-2d-9p-jacobi-1}
  } 
\hfill
\subfloat[2D9P Parallel] {
   \includegraphics[width=0.47\linewidth]{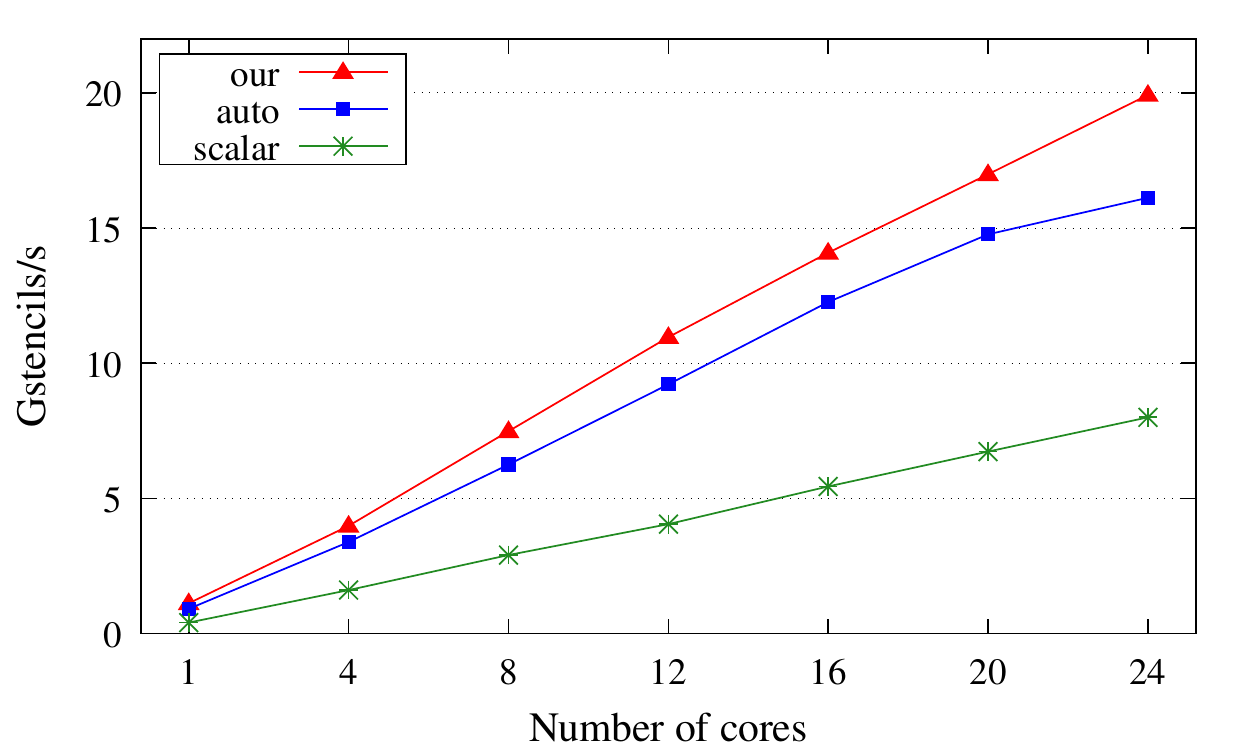}
   \label{fig-2d-9p-jacobi-2}
  }
  \hfill
   \subfloat[Life Sequential] {
   \includegraphics[width=0.47\linewidth]{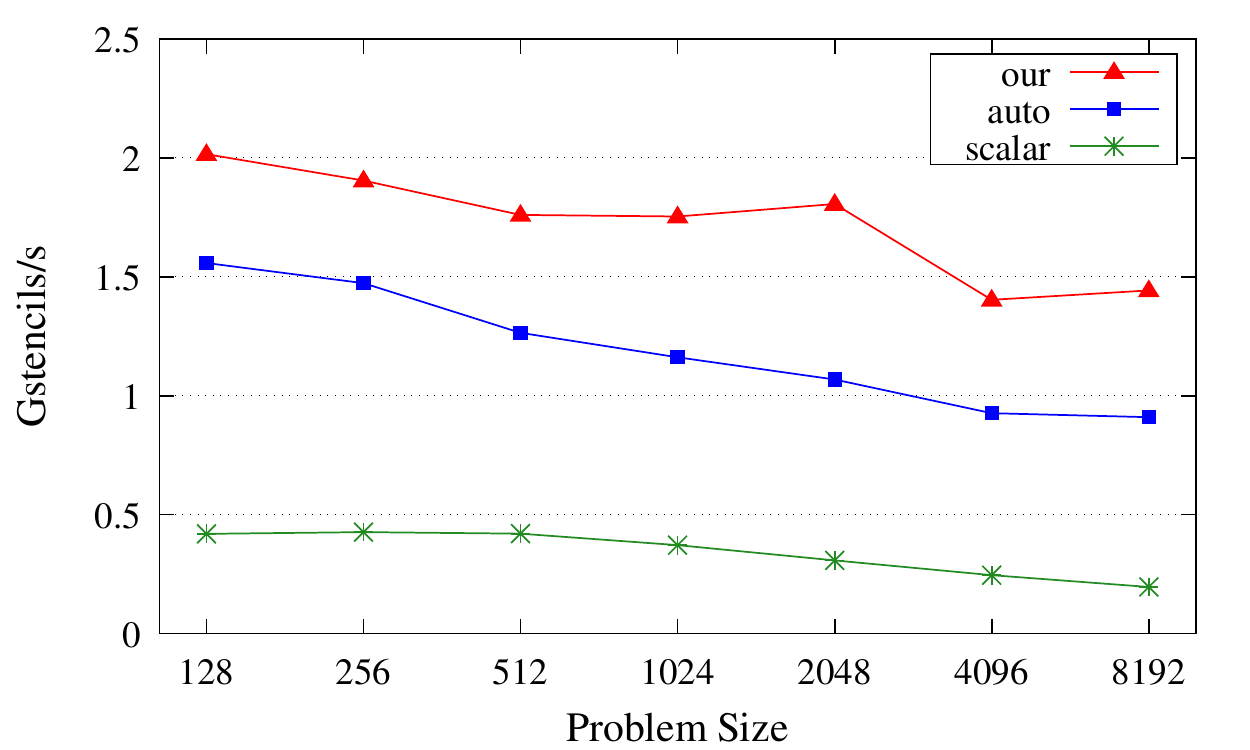}
   \label{fig-2d-life-1}
  } 
\hfill
\subfloat[Life Parallel] {
   \includegraphics[width=0.47\linewidth]{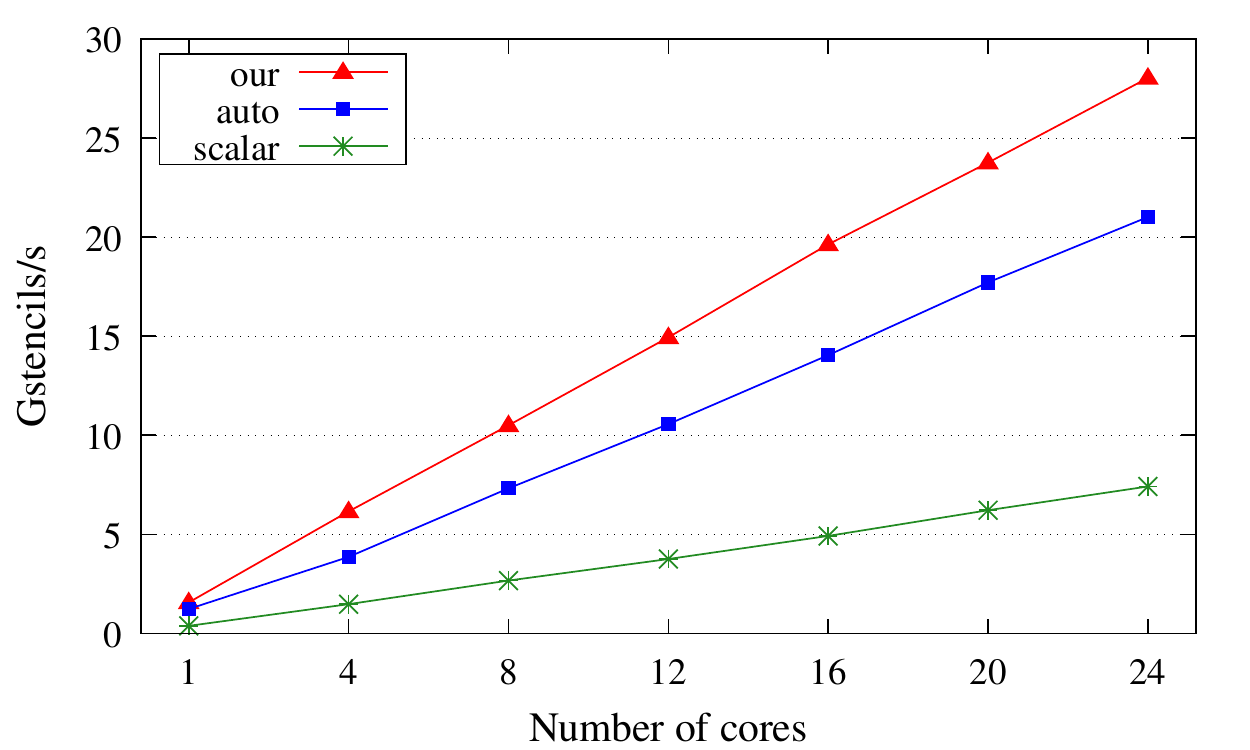}
   \label{fig-2d-life-2}
  }
  \caption{Jacobi Stencils}
\label{fig-jacobi}
\end{figure}

\emph{Jacobi Stencils}. Figure \ref{fig-jacobi} shows 
the performance results of Jacobi stencils.
Left subfigures exhibit the sequential results
and the right subfigures show the parallel performance.

The temporal vectorization of the Heat-1D stencil 
achieves significant performance improvement
over the auto vectorization and scalar code
for problem size larger than 512.
However, the auto-vectorization performs better than the temporal vectorization
for smaller problem sizes.  
The reason is the transpose overhead
of initial and final input vectors assembling
in the temporal vectorization
and this overhead can be amortized
for larger sizes.
The auto-vectorization curve likes a staircase
that contains sharp falls at sizes of cache levels.
This kind of performance curve is ubiquitous. 
For stencil computations, 
this demonstrates that the auto-vectorization
is more sensitive to the cache bandwidth than the scalar code.
On the contrary, the temporal vectorization
produces a flatter curve
due to fewer CPU-cache data accesses.
It indicates that the temporal vectorization is 
less sensitive to the cache bandwidth.
For the parallel results,
they all produce similar scalabilities
with speedup around 20x for 24-core over uni-core.
For all scales, the temporal vectorization
is 3x and 1.6x faster than the scalar code and the auto-vectorization,
respectively.

The Heat-2D and 3D stencils are star stencils.
For continuous output vector calculations,
there is no data sharing over outer space loops.
Thus the data alignment conflict only exists in the uni-stride space dimension
and the auto-vectorization makes an optimal vector utilization
for other space dimensions.
The benefits of the temporal vectorization
may be overweighted by its downsides.
For sequential results, the temporal vectorization
still incur flatter curves than the auto-vectorization.
It obtains competitive and worse performance for
sizes smaller than last-level cache compared
with the auto-vectorization for Heat-2D and 3D stencil,
respectively.
The parallel results are consistent with sequential performance. 
The temporal vectorization exhibits better scalabilities
with large parallelism than the auto-vectorization,
especially for Heat-2D.
This again implies the better bandwidth utilization of the temporal vectorization.
As demonstrated by existing work, the 3D7P Jacobi stencil exhibits limited improvements
on new parallelization  \cite{Bandishti+:sc12}
and vectorization schemes \cite{Henretty+:ics13}.
Our results of Heat-3D reveal a similar phenomenon.

The 2D9P and Life stencils are box stencils.
The auto-vectorization leads to data alignment conflicts
on all space dimensions.
Therefore the temporal vectorization
achieves visible and similar improvements for both stencils.
The Life stencil performs more operations than the 2D9P stencil.
Thus the auto-vectorization is less sensitive to cache bandwidth
as shown in the sequential figure.
Furthermore, the scalability of 2D9P stencil is
similar to that of Heat-2D
and they both have a decline in the 24-core execution
for the auto-vectorization.
The temporal vectorization produces better scalabilities
for both stencils due to fewer data accesses.

\begin{figure}[b]
  \centering
   \subfloat[1D Sequential] {
   \includegraphics[width=0.47\linewidth]{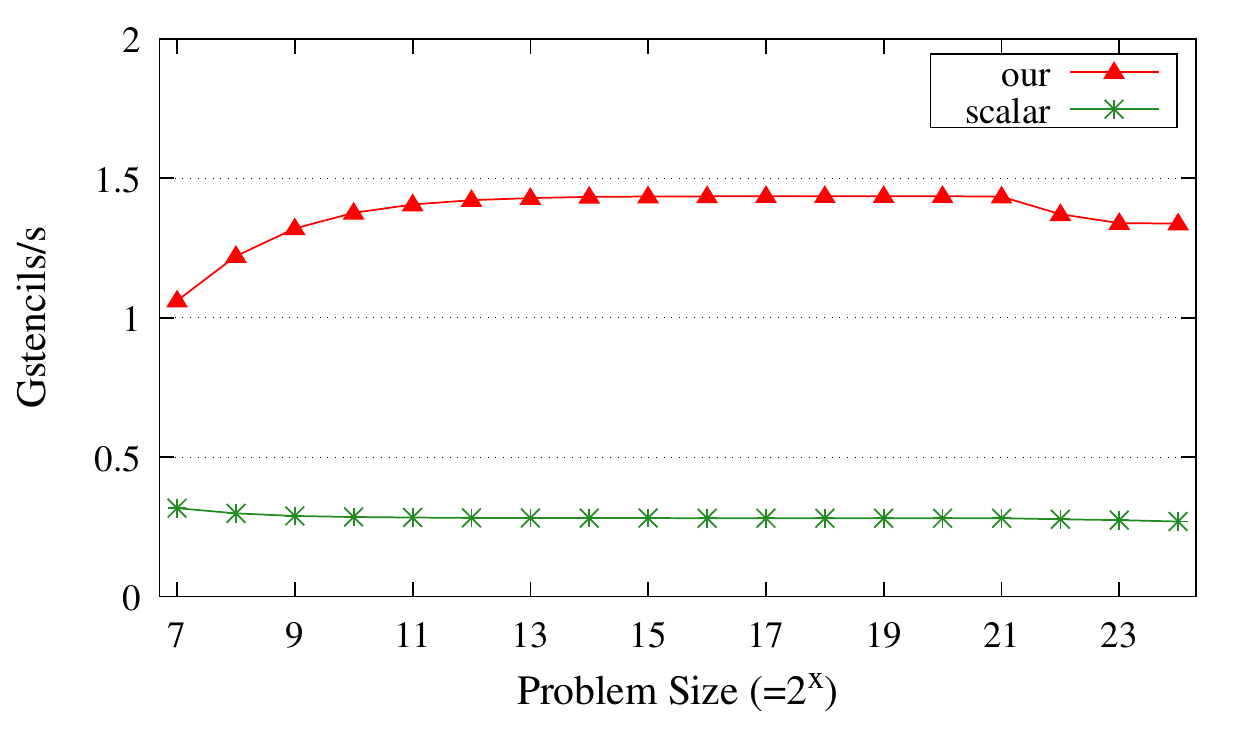}
   \label{fig-1d-gs-3p-1}
  } 
\hfill
\subfloat[1D Parallel] {
   \includegraphics[width=0.47\linewidth]{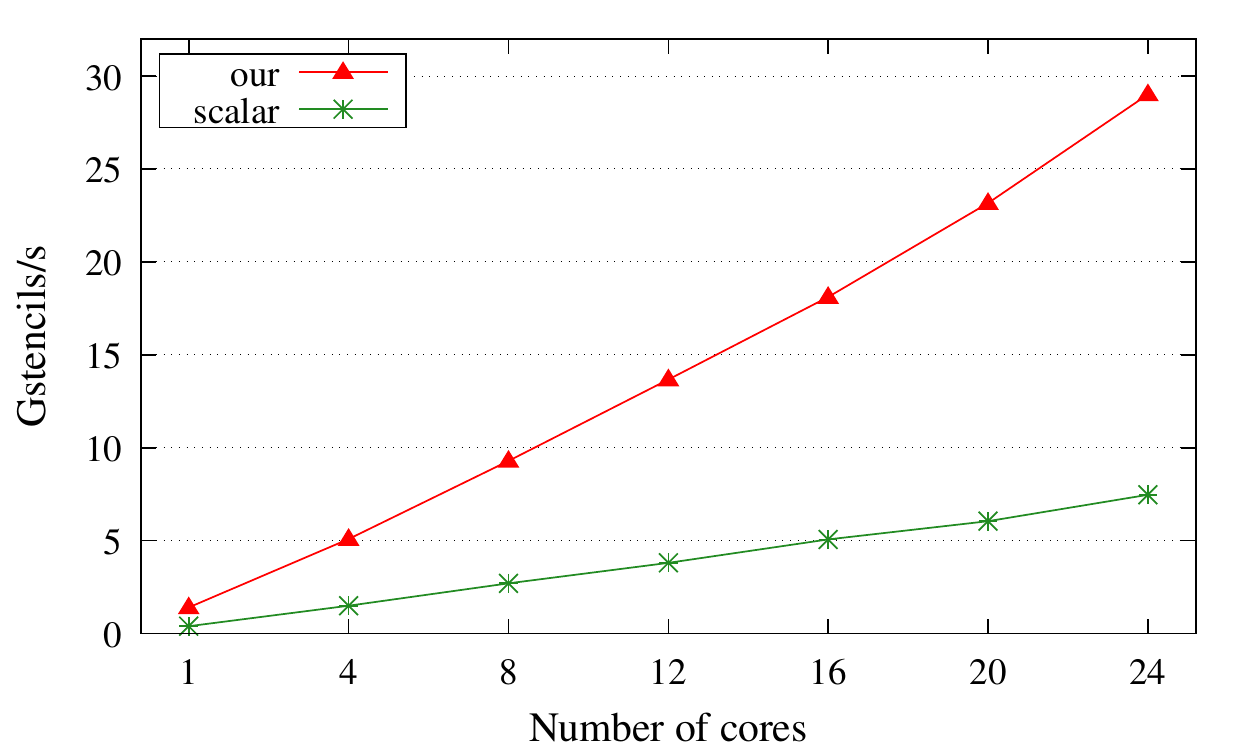}
   \label{fig-1d-gs-3p-2}
  }
 
   \subfloat[2D Sequential] {
   \includegraphics[width=0.47\linewidth]{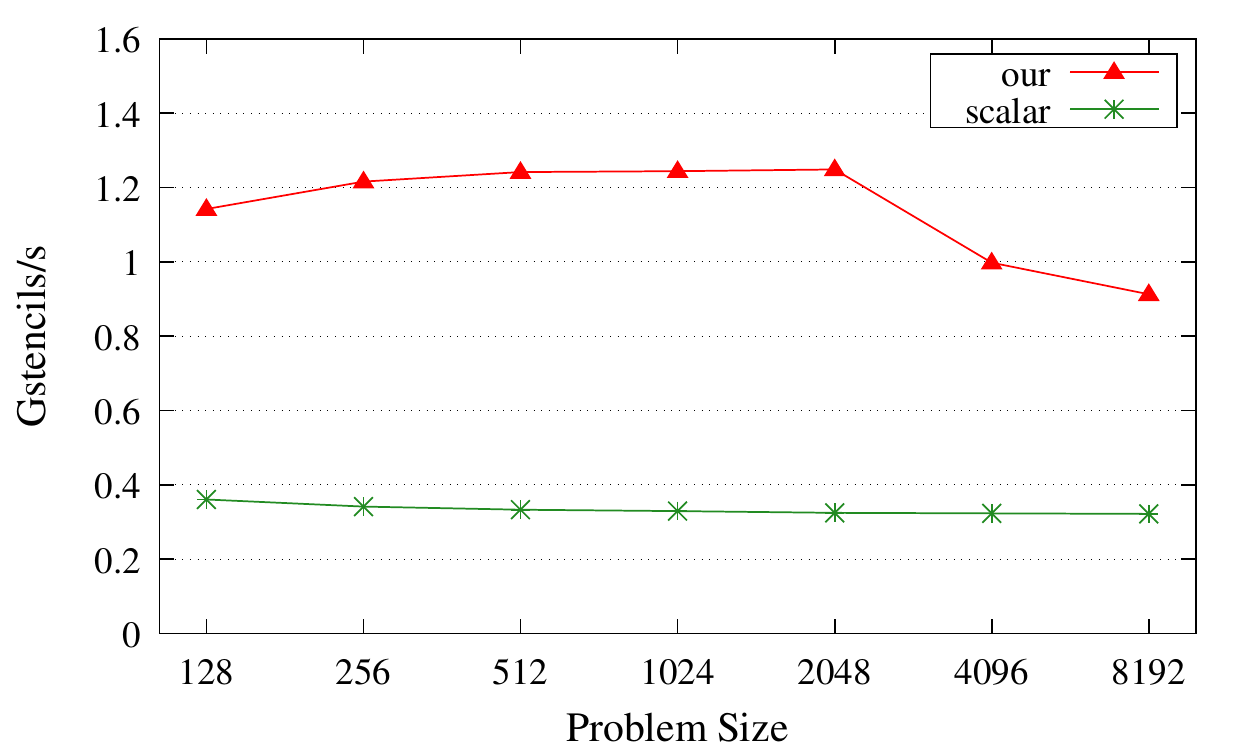}
   \label{fig-2d-gs-5p-1}
  } 
\hfill
\subfloat[2D Parallel] {
   \includegraphics[width=0.47\linewidth]{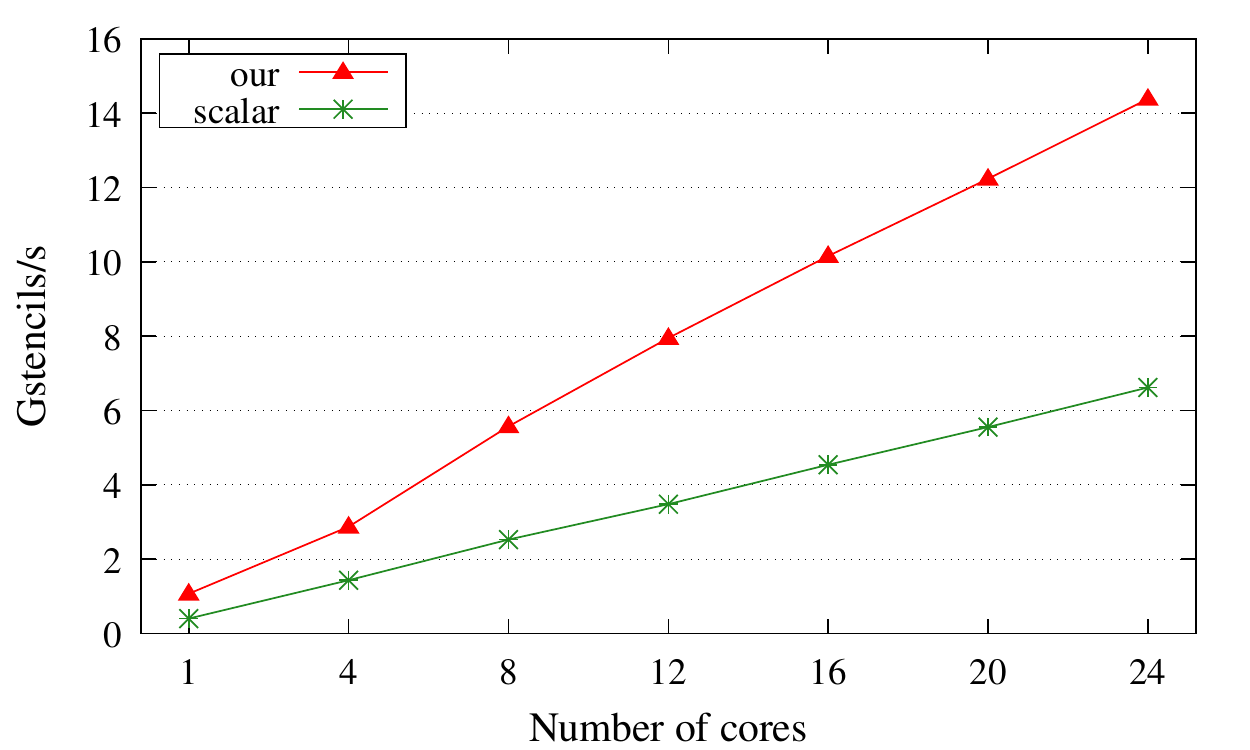}
   \label{fig-2d-gs-5p-2}
  }
 
   \subfloat[3D Sequential] {
   \includegraphics[width=0.47\linewidth]{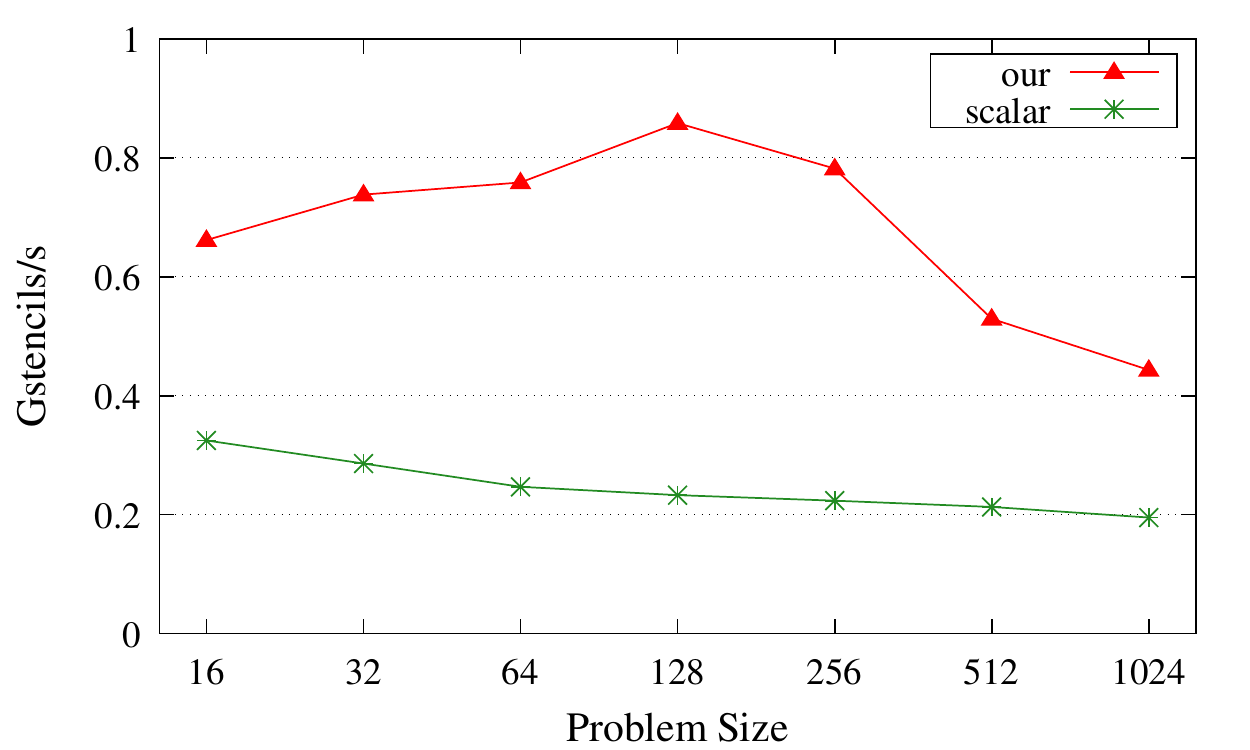}
   \label{fig-3d-gs-7p-1}
  } 
\hfill
\subfloat[3D Parallel] {
   \includegraphics[width=0.47\linewidth]{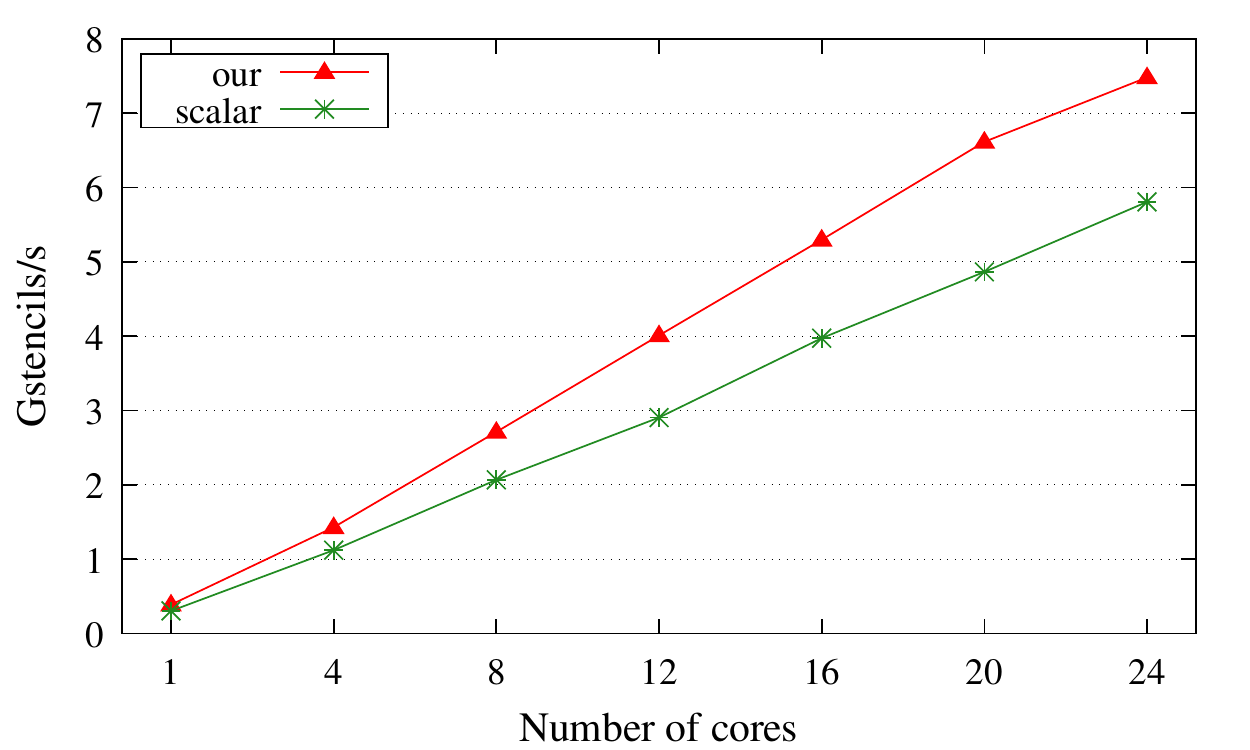}
   \label{fig-3d-gs-7p-2}
  }
  
   \subfloat[LCS Sequential] {
   \includegraphics[width=0.45\linewidth]{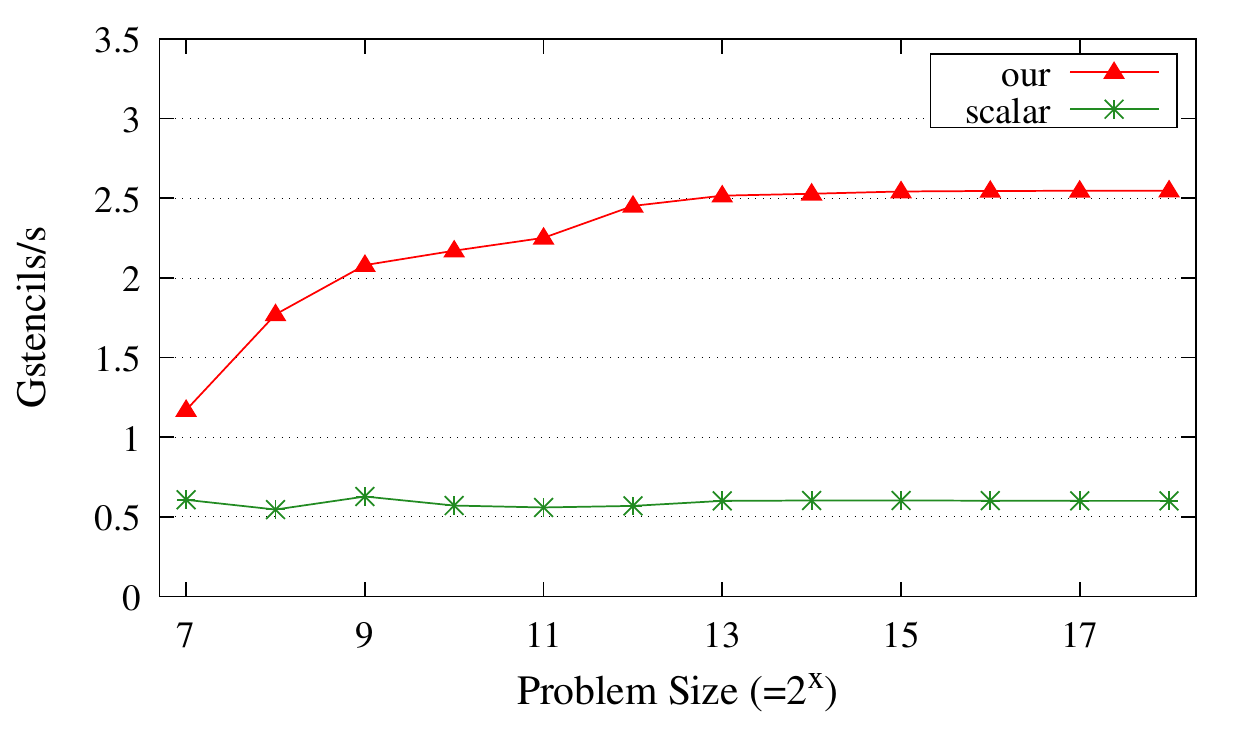}
   \label{fig-lcs-1}
  } 
\hfill
\subfloat[LCS Parallel] {
   \includegraphics[width=0.45\linewidth]{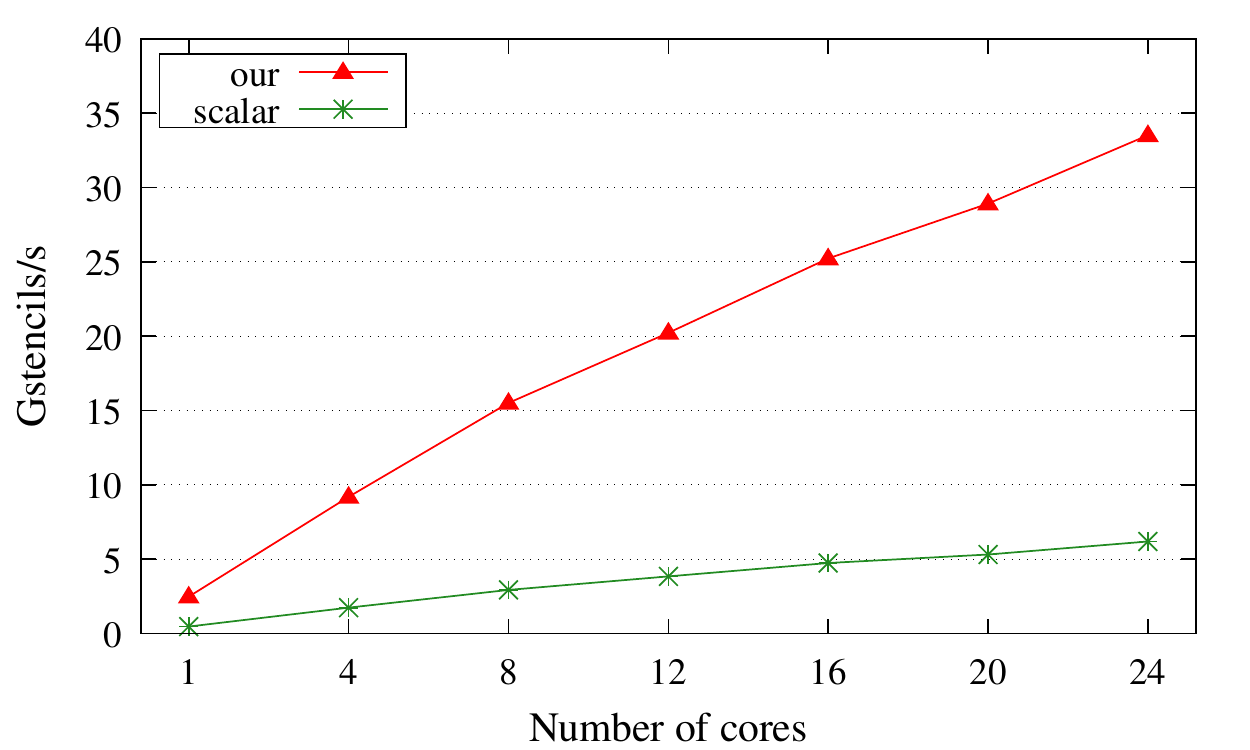}
   \label{fig-lcs-2}
  }
  \caption{Gauss-Seidel Stencils}
\label{fig-gs}
\end{figure}

\emph{Gauss-Seidel Stencils}.
Figure \ref{fig-gs} exhibits the performance results of all Gauss-Seidel stencils.
The temporal vectorization achieves significant performance improvements
for all Gauss-Seidel stencils over the scalar codes.
All sequential executions of the scalar codes without blocking
achieve a similar absolute performance of around $0.4$ Gstencils/s.
This demonstrates that Gauss-Seidel stencils are limited by data dependencies.

For the 1D sequential execution, it obtains a super-linear speedup of up to $4.4$.
The reason is that the temporal vectorization leads to better utilization of the memory bandwidth.
The curve is similar to that of the Head-1D Jacobi stencil.
For smaller problem size
the \emph{scalar ratio}, i.e. the percentage of points processed with 
scalar arithmetic operations
is larger.
For example, with the space stride $s=7$, there are 
$84$ points in a time tile need to be calculated
by the scalar codes (Lines 2-4 and 19-22 in Algorithm \ref{alg-temporalvectorization}).
This leads to a scalar ratio of $16\%$ for problem size $NX=128$
and $1\%$ for $NX=2048$.

For the parallel execution, both scalar and vectorization codes
achieve good scalabilities.
Compared with the single-core execution, the 24-core results reach speedup of $19.6$ for the scalar code and $20.7$
for the vectorization code respectively.
These results are comparable to those of Jacobi stencils
and demonstrate that the parallelogram tiling provides competitive scalabilities.
The temporal vectorization
delivers an average speedup of $3.5$ and a maximal speedup of $3.9$
over the scalar code.

The LCS stencil result is similar to that of the 1D Gauss-Seidel stencil.
It processes integer values with integer SIMD instructions and
has a theoretical maximal speedup of $8$.
The scalar code performs either
$max(lcs[x-1][y]$, $lcs[x][y-1])$ or $lcs[x-1][y-1] + 1$
for calculating $lcs[x][y]$ depending on the equality of $A[x]$ and $B[y]$.
However, the temporal vectorization needs to execute all these computations
and obtains the correct vector by a blend instruction with a mask vector of equalities.
Thus we would expect smaller speedups compared with the 1D Gauss-Seidel stencil.
Nevertheless, the temporal vectorization still achieves good improvements and
yields an average speedup of $5.3$ 
over the scalar code for the parallel execution.

For higher-dimensional Gauss-Seidel stencils,
we see similar trends in sequential results
but more sharp performance decreases when the problem size
is larger than the L3 cache size.
However, the in-cache performance is
relatively steady and it demonstrates
the less sensitivity to the cache bandwidth
of the temporal vectorization.
The maximal speedup is $3.8$ for the 2D stencil with a problem size $2048^2$
and $3.6$ for the 3D stencil with a problem size $128^3$,
respectively.

For the parallel experiments,
both scalar and vectorization codes of the 2D and 3D stencil achieve 
good scalabilities,
e.g. the speedup of 24-core over single core is $16.5$ for scalar and $14$ for vectorization.
Compared with the scalar code, the temporal vectorization obtains
an average speedup of $2.2$ and $1.2$ for 2D and 3D stencils, respectively.









\section{Related Work}
\label{section-relatedwork}


The compiler community has been studying sophisticated universal vectorization techniques
\cite{Allen.Kennedy:toplas87,Allen.Kennedy:book01,Sreraman.Govindarajan:ijpp00,Larsen.Amarasinghe:pldi10,Hampton.Asanovic:cgo08,Nuzman.Zaks:pact08}.
Previous work \cite{Eichenberger+:pldi04,Larsen+:pact02,Wu+:cgo05}
has proposed many solutions to address unaligned vector transfers.
There are also many studies focusing on reducing the data preparation overhead 
\cite{Zhou.Xue:cgo16,Ren+:pldi06,Henretty+:cc11}.
The DLT \cite{Henretty+:cc11} assembles points in the unit-stride space dimension that are free of intra-vector read-read dependencies
in one vector.
Other more general data organization Optimizations
include 
data alignment optimization \cite{Bik+:ijpp02} and 
data interleaving \cite{Nuzman+:pldi06, Zhou.Xue:cgo16},
However, some of these works are too general to capture the specific properties of stencils.
For example, the data alignment conflicts for stencil refer to overlapped vector loads,
it is unable to attack this problem by stream shifts \cite{Eichenberger+:pldi04}.

The state-of-art compilers usually vectorize the innermost loop. 
Outer-loop vectorization techniques \cite{Nuzman.Zaks:pact08} often focus on the case where the innermost loop
is illegal to be vectorized. 
One relaxed legality condition of the outer-loop vectorization is that it is interchangeable with the inner loop. 
However, the time loop is not interchangeable with space loops according to the data dependencies. 
Thought more strict conditions exist, direct outer-loop vectorization at the time loop is still illegal. 
For slightly complicated codes, outer-loop vectorization requires auxiliary arrays 
and programmers need to explicitly declares them \cite{Satish+:isca12}.
Some outer-loop vectorization techniques require the iteration number of the inner loop is invariant with respect to the outer-loop. However, many blocked stencil algorithms 
\cite{Frigo.Strumpen:ics05, Tang+:spaa11,Henretty+:ics13,Yuan+:sc17} 
shrink or expand the data range in all space dimensions as the time proceeds. 
For example, the classic diamond tiling \cite{Krishnamoorthy+:pldi07} 
introduces a diamond shape in the iteration space for a one-dimensional stencil. 
It enlarges the data space in the first half time and then decreases it in the rest time.


There exists a lot of work on improving the register reuse
by utilizing the associative property of stencil calculations.
The fundamental idea is to find a better order of statements across iterations.
Deitz et al \cite{Deitz+:ics01} proposes a compiler formulation and transformation called array common subexpression elimination.
A similar idea called partial sum is also studied in \cite{Basu+:ipdps15}.
Cruz and Araya-polo \cite{Cruz.Araya-polo:toms14} and Stock et al \cite{Stock+:pldi14}
combine the gather (update an output element with all its neighbors)
and scatter (update all its neighbors with one input element) patterns
in one and many space dimensions, respectively.
Several studies \cite{Jin+:cf16,Rawat+:ppopp18,Ma+:jcst16,Zhao+:sc19} describe register reuse frameworks for GPUs.
Yount \cite{Yount:hpcc15}
proposed a vector folding method to group points in the entire data space rather than
a single dimension.
However, prior work either targets scalar registers for high-order stencils on GPUs or  
specific stencils with constant and symmetrical coefficients.
They only consider the reordering in one time step and
data alignment conflicts are inevitable with vectorization. 
Furthermore, these methods are not applicable to and examined for Gauss-Seidel stencils.
The temporal vectorization preserves the same calculation order to the scalar code.
Thus we believe this work can be incorporated into our scheme.

\section{Conclusion}
\label{section-conclusion}
We have presented a new temporal vectorization scheme.
It vectorizes the stencil computation in the whole iteration space and assembles
points with different time coordinate in one vector.
The temporal vectorization leads to a small fixed number of vector reorganizations
that is irrelevant to the vector length, stencil order, and dimension.
Furthermore, it is also applicable to Gauss-Seidel stencils.
The effectiveness of the temporal vectorization is demonstrated
by various Jacobi and Gauss-Seidel stencils.
Future work will design a framework to automatically generate the stencil codes
We will also design an auto-tuning method to efficiently
search the best block size.


\end{document}